\documentclass[preprint,3p,times]{elsarticle}
\usepackage{ecrc}
\usepackage{float}

\volume{00}

\firstpage{1}

\journalname{Physica D}

\runauth{Adriano \& Susanto}

\jid{procs}

\jnltitlelogo{Physica D}

\CopyrightLine{2022}{Published by Elsevier Ltd.}

\usepackage{amssymb}
\usepackage{amsthm}
\usepackage{amsmath}

\usepackage[figuresright]{rotating}

\usepackage{caption}
\usepackage{subcaption}

\usepackage{graphicx}
\usepackage{dcolumn}
\usepackage{bm}
\usepackage{epstopdf}
\usepackage{epsfig}
\usepackage[colorlinks = true,
linkcolor = blue,
urlcolor  = blue,
citecolor = blue,
anchorcolor = blue]{hyperref}
\usepackage{url}

\newcommand{\Z}{\mathbb{Z}}

\newtheorem{theorem}{Theorem}
\newtheorem{lemma}{Lemma}

\begin{document}
	
	\begin{frontmatter}
		\dochead{}
		
		\title{Exponential asymptotics of quantum droplets and bubbles
        }
		
		
		\author{Farrell Theodore Adriano}
		\author{Hadi\ Susanto\corref{cor1}}
		\ead{hadi.susanto@yandex.com}
		
    \address{Department of Mathematics, Khalifa University, PO Box 127788, Abu Dhabi, United Arab Emirates}
		\cortext[cor1]{Corresponding author}

		\begin{abstract}
            This research investigates the formation and stability of localized states, known as quantum droplets and bubbles, in the quadratic-cubic discrete nonlinear Schr\"odinger equation. Near a Maxwell point, these states emerge from two fronts connecting the bistable equilibria. By adjusting a control parameter, we identify a "pinning region" where multiple stable states coexist and are interconnected through homoclinic snaking. We analyze the system's behavior to uncover the underlying mechanisms under strong coupling conditions. Using exponential asymptotics, we determine the pinning region’s width and its dependence on coupling strength, revealing an exponentially small relationship between them. Additionally, we employ eigenvalue counting to establish the stability of these states by computing the critical eigenvalue of their corresponding linearization operator, proving onsite fronts unstable and intersite fronts stable. These theoretical results are validated through numerical simulations, which show excellent agreement with our analytical predictions.
		\end{abstract}
        

		
	\end{frontmatter}
	
	
	\section{Introduction}

One-dimensional solitons of the nonlinear Schr\"odinger (NLS) equation are stable localized wave packets and represent the ground state of the model~\cite{novikov1984theory}. In contrast, multidimensional solitons exhibit fundamentally different behavior, encompassing both fundamental and higher-order states with topological structures. The typical cubic self-attractive nonlinearity in higher dimensions generally renders solitons unstable~\cite{fibich2015nonlinear,kartashov2019frontiers}. This instability arises from critical and supercritical collapse, leading to singular solutions through catastrophic wave collapse (self-compression)~\cite{sulem2007nonlinear}. Consequently, significant efforts have been directed toward identifying physically relevant conditions to stabilize both fundamental and topologically structured multidimensional self-trapped localized modes. These modes exhibit unique features absent in the one-dimensional case, such as higher-dimensional self-trapped states with topological charge (intrinsic vorticity)~\cite{malomed2019vortex}.

A promising theoretical solution to this problem was proposed in~\cite{petrov2015quantum}, which leverages quantum fluctuations as a correction to the mean-field behavior of Bose-Einstein condensates (BECs), as initially predicted by Lee, Huang, and Yang~\cite{lee1957eigenvalues}. This concept has since been experimentally realized in quantum droplets, magnetic quantum gases, and dipolar supersolids, as reviewed in~\cite{chomaz2022dipolar,bottcher2020new}. The Lee-Huang-Yang (LHY) effect manifests as local quartic self-repulsive terms in the corresponding NLS equations under the mean-field approximation. This effect stabilizes localized states that would otherwise collapse by balancing mean-field attraction with LHY repulsion. The resulting superfluid state has a density below the maximum limit, rendering it incompressible. This quantum macroscopic state, termed "quantum droplets," represents a novel fluid-like phase~\cite{luo2021new}.
 
In BECs with cigar-like or pancake-like geometries, which arise in a three-dimensional context, these systems can be effectively modeled using one-dimensional or two-dimensional NLS equations. When a strong external potential imposes confinement in one or more spatial directions, the effective NLS equation can be modified, enabling the formation of droplets in lower dimensions. In the one-dimensional limit, the LHY correction introduces a quadratic self-attraction, contrasting with its quartic behavior in the original three-dimensional framework~\cite{petrov2016ultradilute}.

The behavior and properties of one-dimensional quantum droplets were explored in~\cite{astrakharchik2018dynamics}, where solutions to the modified Gross-Pitaevskii equation revealed two distinct physical regimes: tiny droplets with Gaussian-like profiles and large puddles with broad, flat-top structures. Furthermore, Zhou et al.~\cite{zhou2019dynamics} investigated these droplets under a linear optical lattice, uncovering phenomena such as multistability and shifts in stability between solitons located at or between lattice sites. Subsequent studies demonstrated that these multi-stable droplets exhibit multi-hysteresis behavior~\cite{dong2020multi}, existing within specific intervals of the band-gap spectrum. In another study, Kartashov and Zezyulin~\cite{kartashov2024enhanced} analyzed bright solitons in an optical lattice, finding that they display enhanced mobility with alternating patterns of mobility and immobility related to their multistability. Similarly, Zhao et al.~\cite{zhao2021discrete} examined deep lattice scenarios, where the NLS equation reduces to a discrete model, showing that multistability also emerges in this regime.

The multistability observed in~\cite{zhou2019dynamics,dong2020multi,zhao2021discrete} is a manifestation of the pinning phenomenon, which is common in various nonlinear dynamical systems where two or more uniform states are bistable (see, for example,~\cite{susanto2011variational,matthews2011variational} and references therein). In such systems, a front connecting the states typically shifts depending on which state is energetically favorable. However, at a specific parameter value, known as the \emph{Maxwell point}, neither state is favored, resulting in a stationary front. When two such fronts are placed back-to-back, they form a localized structure. The bifurcation diagram for this localized solution, plotted against a control parameter, exhibits a distinctive snaking pattern characterized by a series of turning points that closely align across a range of parameters near the Maxwell point. This defines a \emph{pinning region}, and since the localized state transitions from and returns to a uniform state, this phenomenon is referred to as \emph{homoclinic snaking}~\cite{woods1999heteroclinic}.

Recently, the present authors and their collaborators \cite{kusdiantara2024analysis} studied the homoclinic snaking and pinning regions of discrete quantum droplets and bubbles. The width of the pinning region was analyzed in the uncoupled and continuum limits, where in those limits, the width depends algebraically and exponentially on the coupling constant, respectively. The exponential asymptotic analysis was done using a variational approximation \cite{susanto2011variational,matthews2011variational}, where they presented the pinning region's width estimate. 

In this work, we study the multistability and the pinning region of quantum droplets and bubbles using exponential asymptotics beyond all orders. While we obtain the same exponentially small scaling of the pinning region width estimate, our current approach is tractable mathematically and provides better agreement. Moreover, we establish the stability of discrete droplets and bubbles analytically, which was absent in the previous studies. Their critical eigenvalues are also compared with the numerics, where we obtain excellent agreement. 

Exponential asymptotics, or beyond-all-orders asymptotics, is a powerful method that extends traditional perturbation techniques by including exponentially small contributions often ignored in classical asymptotic expansions \cite{boyd2012weakly}. Stokes line analysis \cite{stokes1851numerical,stokes1864discontinuity} is vital to track how exponentially small terms in the complex plane switch on or off as parameters vary. Kozyreff and Chapman \cite{kozyreff2006asymptotics,chapman2009exponential} initiated the use of exponential asymptotics in the study of homoclinic snaking, where they demonstrated that these exponentially small terms are significant in the interaction between localized structures. They uncovered phenomena such as front interactions and pinning effects, which were crucial to the detailed understanding of homoclinic snaking and governed the intricate snaking behavior. Leading-order asymptotics do not capture these contributions and require unique resolution methods. Recent work has extended the use of exponential asymptotics to handle problems in homoclinic snaking \cite{dean2011exponential,dean2015orientation,de2019beyond}.

This paper is organized as follows. Section~\ref{sec2} introduces the model and its discrete quantum droplets, which are localized states. In Section~\ref{sec3}, we develop and use exponential asymptotics to construct the droplets in the continuum limit, where we show two fundamental localized states, i.e., intersite and onsite. We subsequently analyze the snaking mechanism. We then provide numerical results and compare them with our analysis. We prove the stability of the quantum droplets in Section~\ref{sec4}. Finally, Section~\ref{sec5} summarizes our findings and concludes the paper.

\section{Mathematical model}\label{sec2}

 \begin{figure}[tbhp]
     \centering
     \begin{subfigure}{1\textwidth}
         \includegraphics[width = \textwidth]{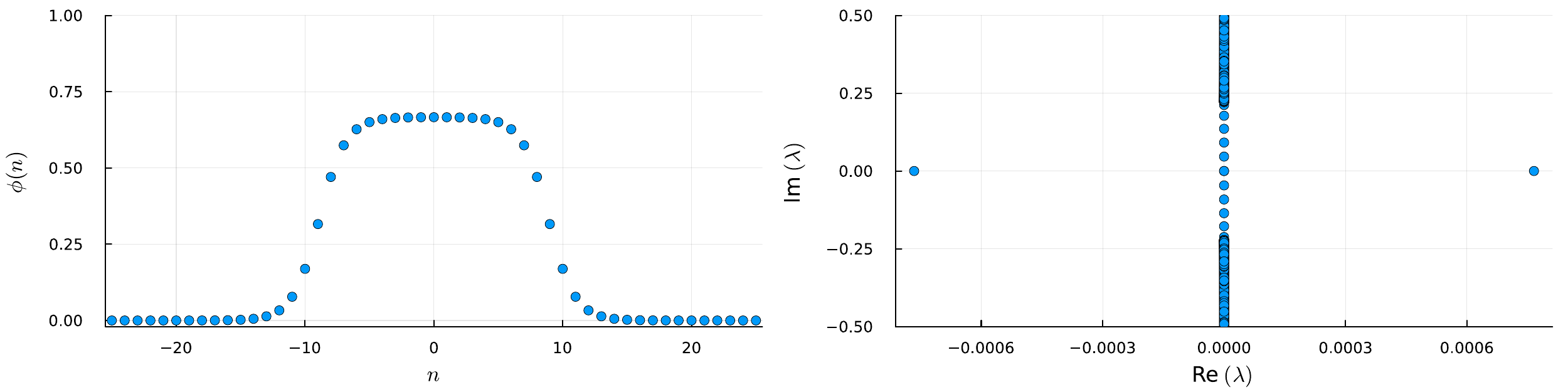}
         \caption{}
     \end{subfigure}
     \begin{subfigure}{1\textwidth}
         \includegraphics[width = \textwidth]{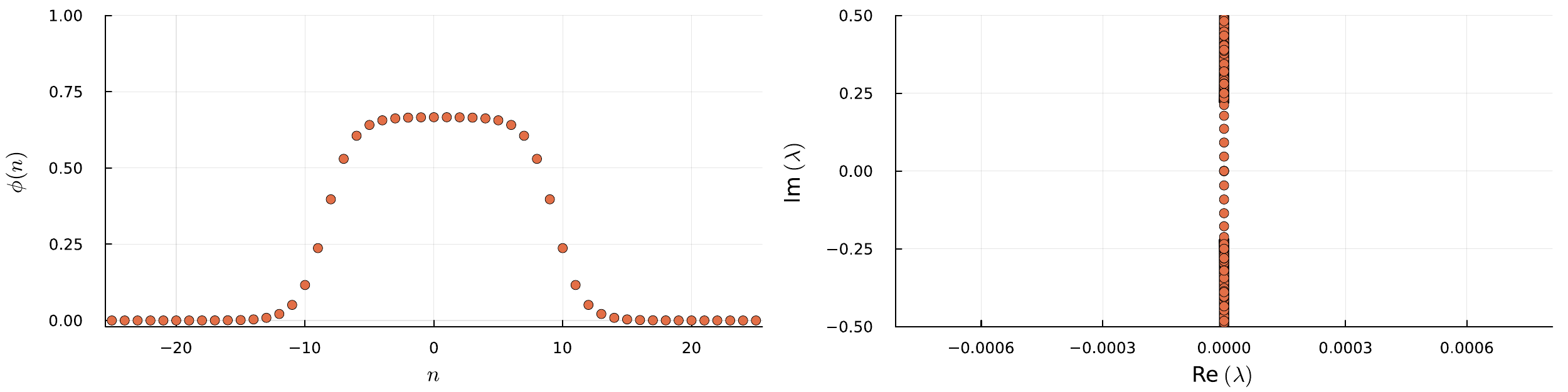}
         \caption{}
     \end{subfigure}
     \begin{subfigure}{1\textwidth}
         \includegraphics[width = \textwidth]{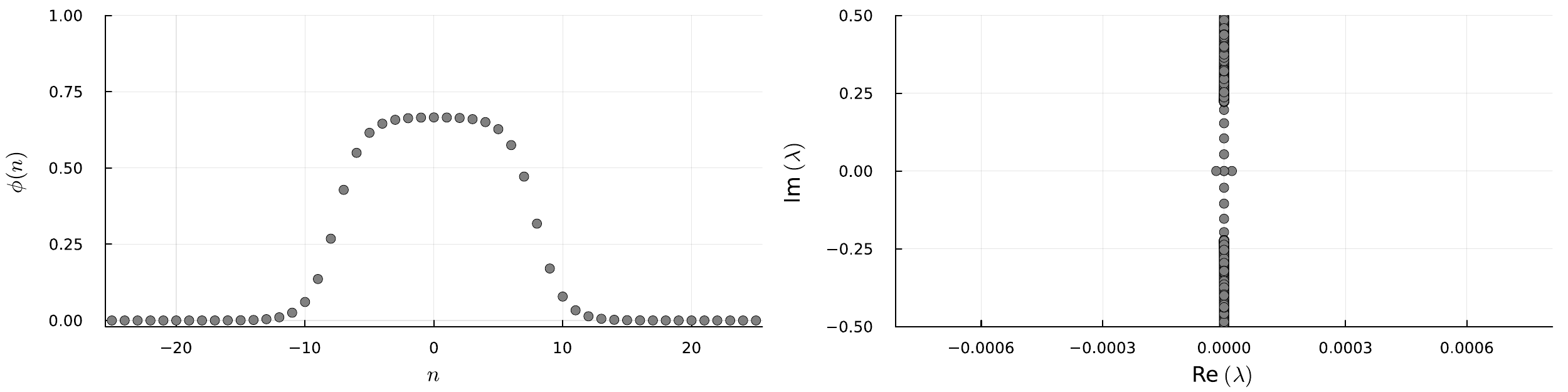}
         \caption{}
     \end{subfigure}
     \caption{Profiles (left) and spectrum (right) of discrete droplets for \(C = 0.5\). The presence of a spectrum with nonzero real part indicates instability. Panel (a) shows an onsite droplet, which is symmetric about a lattice site (here, \(n = 0\)); panel (b) shows an intersite droplet, symmetric about the midpoint between two lattice sites (\(n = 1/2\)); and panel (c) shows a ladder droplet solution.
     }
     \label{fig:droplets_profile_spectrum}
 \end{figure}

We consider the following dimensionless discrete NLS equation with quadratic and cubic nonlinearity
\begin{equation} \label{eqn:dnls_main}
    i\partial_{t}\phi = -\dfrac{C}{2}\Delta \phi + \left[-\mu \phi - |\phi|\phi + |\phi|^{2}\phi\right]
\end{equation}
where $\phi = \phi(n)$, $n \in \Z$, is the wave packet of the quantum droplet, $C$ denotes the coupling strength between lattice sites, and $\mu$ denotes a chemical potential which will be treated as a bifurcation parameter. In the physical context of Bose-Einstein condensates, the defocusing cubic term represents the mean-field two-body interactions between particles. While mean-field theory treats interactions in a classical approximation, it neglects the intrinsic quantum "jiggles" present in the system. The LHY correction accounts for the energy contribution from these fluctuations, specifically the zero-point energy associated with collective excitations (Bogoliubov modes) in the condensate. The attractive quadratic term represents this leading-order correction and is often referred to as the LHY term \cite{zhou2019dynamics,luo2021new}.

The mathematical form of the LHY correction in the NLS equation depends sensitively on the dimensionality of the system. In one dimension, the LHY energy density scales as $n^{3/2}$ \cite{petrov2016ultradilute}, where $n = |\psi|^2$ denotes the particle density, leading to a correction term $n^{1/2} \psi = |\psi| \psi$ in \eqref{eqn:dnls_main} above. In two dimensions, the energy density is proportional to $n^2 \ln(n)$ \cite{petrov2016ultradilute}; although more complex, this can be approximated by $n^2$ in certain regimes, yielding a cubic nonlinearity of the form $|\psi|^2 \psi$. In three dimensions, the energy density scales as $n^{5/2}$ \cite{petrov2015quantum}, producing a term $n^{3/2} \psi = |\psi|^3 \psi$, which corresponds to a quintic nonlinearity.

Equation \eqref{eqn:dnls_main} has a Hamiltonian/potential, which is given by
\begin{equation}
    E(\phi) = \sum_{n = -\infty}^{\infty}\left[\dfrac{C}{2}|\phi(n)-\phi(n-1)|^{2} + \dfrac{\mu}{2}|\phi(n)|^{2} + \dfrac{1}{3}|\phi(n)|^{3} - \dfrac{1}{4}|\phi(n)|^{4}\right].
\end{equation}
The quadratic and cubic nonlinearities considered in our model belong to a power-law nonlinearity of the form \( |\psi|^{2\sigma} \psi \) with exponent \( \sigma = 1/2,1 \). They lie within the subcritical regime for the \emph{continuous} NLS equation in one dimension. It is well known that, for such subcritical nonlinearities, the Cauchy problem is globally well-posed in the Sobolev space \( H^1 \) \cite{fibich2015nonlinear}. In this setting, the conservation of mass and energy prevents finite-time blow-up, ensuring that solutions with initial data in \( H^1 \) exist for all time. In the \emph{discrete} NLS equation, the Cauchy problem is believed to be globally well-posed for any $\sigma>0$ \cite{laedke1994stability,tzirakis2005collapse}.

We will study the stationary solutions to \eqref{eqn:dnls_main} in the continuum limit, i.e. $C \to \infty$. Seeking real stationary solutions, we will consider the stationary equation 
\begin{equation}\label{eqn:dnls_stationary_real}
    0 = \frac{C}{2}\Delta \phi - (-\mu \phi - |\phi|\phi + \phi^{3}).
\end{equation}
To analyze near the continuum limit $C \to \infty$, we write $C = 2/{\varepsilon^{2}}$, and thus \eqref{eqn:dnls_stationary_real} is equivalent to
\begin{equation}\label{eqn:dnls_stationary_real_eps}
    0 = \Delta \phi - \varepsilon^{2}(-\mu \phi - |\phi|\phi + \phi^{3}),
\end{equation}
where $\varepsilon \to 0$. For brevity, we will write
\[
F(\phi;\mu) = -\mu \phi - |\phi|\phi + \phi^{3}.
\]

Soliton solutions of \eqref{eqn:dnls_main} can be obtained by seeking homoclinic or heteroclinic trajectories of the stationary equation \eqref{eqn:dnls_stationary_real}. These trajectories, if they exist, depend on the uniform (homogeneous) solutions of \eqref{eqn:dnls_stationary_real}, which are given by:
\begin{equation}
    \phi^{(0)} = 0, \quad \phi^{(1)} = \dfrac{1-\sqrt{4\mu+1}}{2}, \quad \phi^{(2)} = \dfrac{1+\sqrt{4\mu+1}}{2}.
\end{equation}
Linear stability analysis reveals that the uniform solutions of \eqref{eqn:dnls_main} exhibit bistability between $\phi^{(0)}$ and $\phi^{(2)}$, which holds for all values of $\mu$ admitting real uniform solutions, i.e. $\mu > -1/4$. Moreover, there exists a critical value of $\mu$ such that the potential at these two uniform solutions vanishes simultaneously, making the two solutions energetically similar. This occurs at the Maxwell point $\mu = \mu_{M} = -2/9$.

The Laplacian term in the continuum limit $C \to \infty$ in \eqref{eqn:dnls_stationary_real} will become a second derivative. In this limit, $\mu = \mu_{M}$ implies the existence of a heteroclinic trajectory connecting $\phi^{(0)}$ and $\phi^{(2)}$, corresponding to a stationary front. When $\mu > \mu_{M}$, $\phi^{(0)}$ has a lower potential than $\phi^{(2)}$, making it energetically favorable. This results in a homoclinic trajectory emanating from $\phi^{(0)}$ called a droplet. Conversely, when $\mu < \mu_{M}$, $\phi^{(2)}$ has a lower potential than $\phi^{(0)}$, and the homoclinic trajectory emanates from $\phi^{(2)}$, forming a bubble. See \cite{kusdiantara2024analysis,katsimiga2023interactions,katsimiga2023solitary}.

In the discrete system, localized solutions to \eqref{eqn:dnls_main} exist and represent discrete droplets, as illustrated in Fig.\ \ref{fig:droplets_profile_spectrum}. There are two fundamental types of droplets, distinguished by the location of their center: onsite and intersite \cite{kusdiantara2024analysis}. Onsite droplets are centered at a lattice site, i.e., they are symmetric about some \( n_{0} \in \mathbb{Z} \) where the droplet attains its maximum amplitude. Intersite droplets, by contrast, are centered at the midpoint between two adjacent lattice sites, i.e., they are symmetric about a point \( n_{0} \in \mathbb{Z} + 1/2 \). Both types of droplets exist near the Maxwell point, where the energies of the two uniform states are balanced. In this regime, droplets are formed by the joining of two stationary fronts. In addition to these symmetric structures, there also exist ladder solutions, which are asymmetric, i.e., not necessarily symmetric with respect to a lattice site or the midpoint between two neighboring sites.

\begin{figure}[tbhp]
     \centering
     \includegraphics[width=0.7\linewidth]{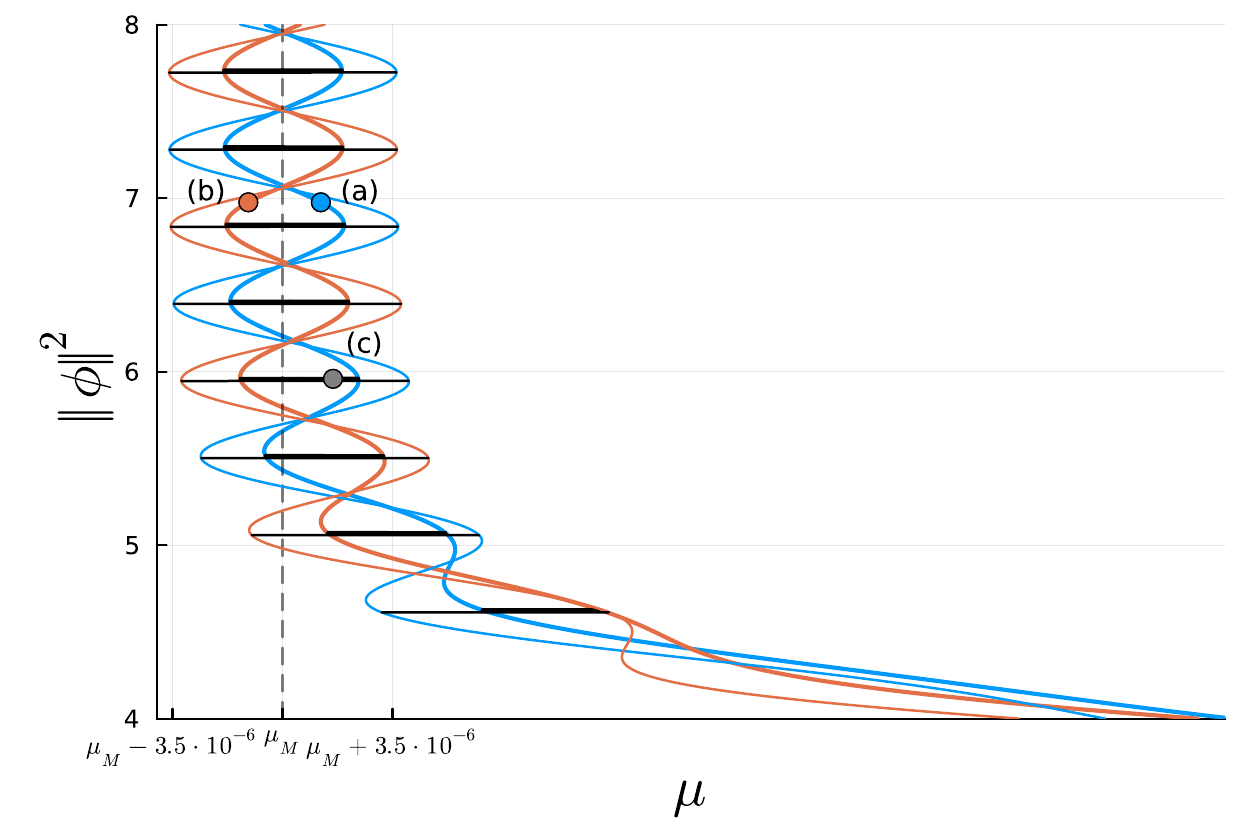}
     \caption{Bifurcation diagram of droplets for $C = 0.5$. The squared $l^{2}(\mathbb{Z})$ norm of the solutions are plotted against the bifurcation parameter $\mu$. The thick lines represent the numerically calculated solutions and the thin lines represent the approximation obtained from the asymptotic formulae \eqref{eqn:dmu_snake_asymp} and \eqref{eqn:dmu_ladder_asymp}. The blue, red, and black lines represent the onsite, intersite, and ladder droplet solutions respectively.
     The positions of the profiles in Figure \ref{fig:droplets_profile_spectrum} on the bifurcation diagram are indicated as points (a), (b), and (c).}
     \label{fig:bifurcation_diag_C=0.5}
 \end{figure}

When the coupling parameter \(C \gg 1\), these droplets exhibit snaking behavior at an exponentially small level, as shown in Fig.\ \ref{fig:bifurcation_diag_C=0.5}. Snaking refers to the characteristic pattern in the bifurcation diagram, where the solution branches oscillate back and forth, creating a "snake-like" structure. As the onsite and intersite branches weave through one another in the bifurcation diagram, the ladder solutions connect these two branches, making it appear as ladder rungs between the snakes. The parameter region where this snaking occurs is known as the pinning region, which arises due to the discreteness of the system and the interaction between the fronts.

\begin{figure}[tbhp]
    \centering
    \includegraphics[width=\linewidth]{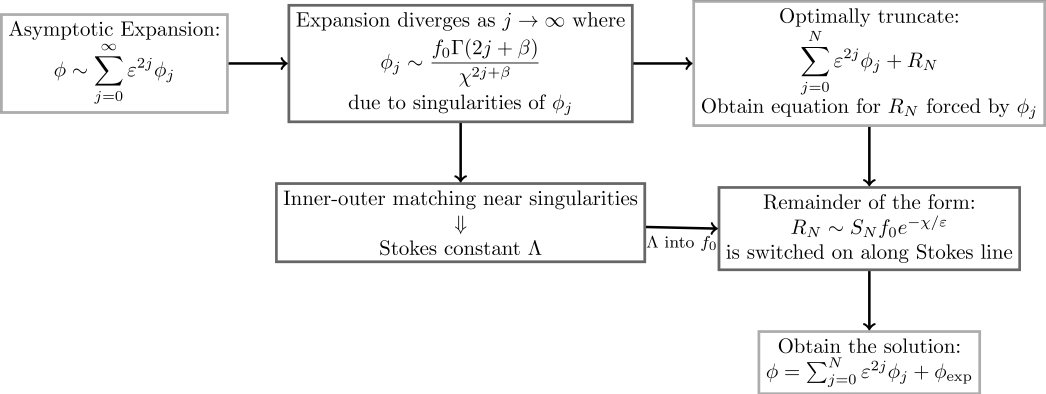}
    \caption{Scheme of the exponential asymptotics procedure in this work.}
    \label{fig:exp_asymp_diag}
\end{figure}

To analyze this snaking phenomenon, we employ exponential asymptotics, a powerful technique for capturing exponentially small effects that are beyond the reach of standard perturbation methods. The general scheme for this approach is illustrated in Fig.\ \ref{fig:exp_asymp_diag}, which outlines the steps involved in deriving the asymptotic behavior of the solutions and identifying the pinning region.

\section{Asymptotic Behavior near Maxwell Point}\label{sec3}

We now analyze the behavior of the stationary front solution (heteroclinic trajectory) of \eqref{eqn:dnls_stationary_real_eps} near the Maxwell point, i.e., when $\mu = \mu_{M} + \delta\mu$ with $|\delta\mu| \ll 1$. To this end, we introduce the slow scale $x = \varepsilon n$ and $x_{0} = \varepsilon n_{0}$, where $n_{0}$ is a shifting constant that describes the pinning mechanism. Defining $\tilde{x} = x - x_{0}$, we express our equations in terms of the shifted variable $\tilde{x}$.

Under the slow variable $\tilde{x}$, the Laplacian $\Delta \phi$ term transforms as follows:
\begin{align*}
    \Delta \phi(\tilde{x}) &= \phi(\tilde{x} + \varepsilon) + \phi(\tilde{x} - \varepsilon) - 2\phi(\tilde{x}) \\
    &= 2\sum_{m \geq 1} \dfrac{\varepsilon^{2m}}{(2m)!}\partial_{\tilde{x}}^{2m} \phi.
\end{align*}
We now expand $\phi(\tilde{x})$ in an asymptotic power series in $\varepsilon$ as
\begin{equation} \label{eqn:expansion_phi}
    \phi(\tilde{x}) \sim \sum_{j=0}^{\infty} \varepsilon^{2j}\phi_{j}(\tilde{x}), \quad \text{as} \quad \varepsilon \to 0.
\end{equation}
Substituting the expansion \eqref{eqn:expansion_phi} into \eqref{eqn:dnls_stationary_real_eps} yields the following equation for $\phi_{0}$ at leading order:
\begin{equation}\label{eqn:phi_0_equation}
    0 = \partial_{\tilde{x}}^{2} \phi_{0}(\tilde{x}) - F(\phi_{0};\mu_{M}).
\end{equation}
Taking $\phi_{0}(\tilde{x})$ to be the heteroclinic trajectory connecting $\phi^{(0)}$ and $\phi^{(2)}$, i.e., 
\[
\lim_{\tilde{x}\to-\infty}\phi_{0} = \phi^{(0)} \quad \text{and} \quad \lim_{\tilde{x}\to\infty}\phi_{0} = \phi^{(2)},
\]
we obtain the leading-order solution:
\begin{equation} \label{eqn:phi_0_soln}
    \phi_{0}(\tilde{x}) = \dfrac{2/3}{1+e^{-\sqrt{2}\tilde{x}/3}}.
\end{equation}
At higher orders $\mathcal{O}(\varepsilon^{2j+2})$, we derive differential equations for $\phi_{j}$. For real $\tilde{x}$, these equations can be solved successively to determine the terms $\phi_{j}$ for each $j$, providing the correction terms of $\phi$ at each algebraic order of $\varepsilon$. However, as we will demonstrate, these algebraic-order corrections are insufficient to capture the pinning phenomenon induced by the discreteness of the problem.

\begin{figure}[tbhp]
    \centering
    \includegraphics[width=0.6\linewidth]{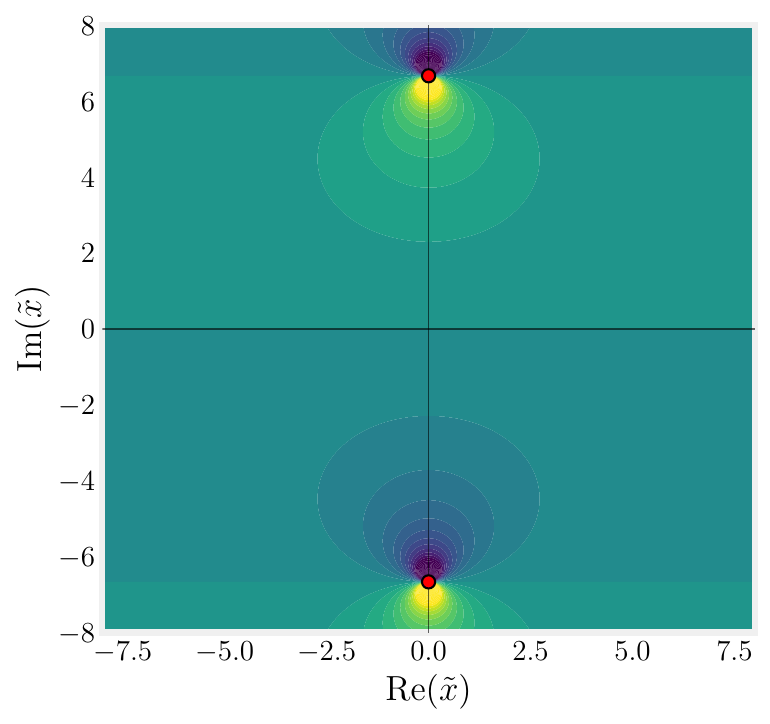}
    \caption{Plot of $\text{Im}(\phi_{0}(\tilde{x}))$ in the complex plane, showing the presence of a pair of singularities.}
    \label{fig:Im_phi0}
\end{figure}

By considering the analytic continuation of $\phi_{0}$ in the complex plane, we observe that $\phi_{0}$ has first-order poles at $\tilde{x} = \zeta =  i \frac{3}{\sqrt{2}}(\pi + 2k\pi)$ for $k \in \mathbb{Z}$; see Fig.\ \ref{fig:Im_phi0}. Near these poles, $\phi_{0}$ diverges as
\begin{equation}
    \phi_{0} \sim \dfrac{\sqrt{2}}{\tilde{x}-\zeta} + \dfrac{1}{3}.
\end{equation}
These singularities become more pronounced in the higher-order terms $\phi_{j}$ as $j \to \infty$. This behavior leads to a Stokes phenomenon, where exponentially small remainder terms emerge in the front solutions of \eqref{eqn:dnls_stationary_real_eps} as a Stokes line is crossed in the complex plane. The remainder term is crucial in determining the pinning region $\delta\mu$ that admits these front solutions.

\subsection{Late Order Terms}

The late-order terms $\phi_{j}$, as $j \to \infty$, are key to understanding the divergent behavior of the series and the remainder term. We therefore examine the asymptotic properties of $\phi_{j}$ for large $j$. At $\mathcal{O}(\varepsilon^{2j+2})$, $\phi_{j}$ satisfies the equation
\begin{equation} \label{eqn:late_order_phi_n}
    0 = 2\sum_{p=1}^{j+1}\dfrac{1}{(2p)!} \partial_{\tilde{x}}^{2p}\phi_{j-p+1} - F_{\phi;M}(\phi_{0})\phi_{j} + \dots,
\end{equation}
where $F_{\phi;M}(\phi_{0}) = F_{\phi}(\phi_{0};\mu_{M})$. It can be verified a posteriori that the terms omitted in \eqref{eqn:late_order_phi_n} are of higher order in $1/j$.

From \eqref{eqn:late_order_phi_n}, we see that $\phi_{j}$ is generated by differentiating $\phi_{j-1}$ four times, $\phi_{j-2}$ six times, and so forth, before integrating twice. This process implies that if $\phi_{j-1}$ has a pole of order $l$ at $\zeta$, then $\phi_{j}$ will inherit a pole at $\zeta$ with order $l+2$. Thus, the poles of $\phi_{j}$ align with those of the leading-order term $\phi_{0}$, but the pole order increases by 2 with each successive $j$. Additionally, the repeated differentiation suggests that $\phi_{j}$ grows factorially with $j$, driven by the singular behavior at these poles.

Based on this observation, we propose the following asymptotic ansatz for $\phi_{j}$, characterized by a factorial-over-power growth:
\begin{align} \label{eqn:late_order_phi_ansatz}
    \phi_{j}(\tilde{x}) \sim (-1)^{j} \dfrac{\Gamma(2j+\beta)}{[\chi(\tilde{x})]^{2j+\beta}}f_{0}(\tilde{x}),\ \  \text{as} \, j \to \infty.
\end{align}
Here, $\chi(\tilde{x})$ denotes the singularity location, vanishing at $\zeta$, and $\beta$ is a constant determined by matching the pole order in $\phi_{j}$.

Substituting the ansatz \eqref{eqn:late_order_phi_ansatz} into \eqref{eqn:late_order_phi_n} leads to, at leading order $\mathcal{O}(\Gamma(2j+\beta+2))$:
\begin{align*}
    0 = \dfrac{(-1)^{j+1}2f_{0}}{\chi^{2j+\beta+2}} \sum_{p=1}^{j+1}\dfrac{(-1)^{p}(\chi')^{2p}}{(2p)!} &\approx \dfrac{(-1)^{j+1}2f_{0}}{\chi^{2j+\beta+2}}\sum_{p=1}^{\infty} \dfrac{(-1)^{p}(\chi')^{2p}}{(2p)!} = \dfrac{(-1)^{j+1}2f_{0}}{\chi^{2j+\beta+2}} (\cos \chi' - 1).
\end{align*}
For $f_{0}$ to remain nontrivial, we require $\cos \chi' - 1 = 0$. This condition implies that $\chi' = \kappa = \kappa_{k} = 2k\pi$ for $k \in \mathbb{Z}$, with $k > 0$ taken without loss of generality. Since $\phi_{j}$ inherits the pole structure of $\phi_{0}$, we conclude that
\begin{equation}\label{eqn:chi}
    \chi = \kappa_{k}(\tilde{x}-\zeta) = 2\pi k (\tilde{x}-\zeta),
\end{equation}
where $k \in \mathbb{Z}$.

By taking higher-order terms up to $\mathcal{O}(\Gamma(2j+\beta))$ in \eqref{eqn:late_order_phi_n}, 
substitution of the ansatz \eqref{eqn:late_order_phi_ansatz} leads to
\begin{align}
    \begin{split}
    0 = \ &\Bigg\{\dfrac{(-1)^{j+2}}{\chi^{2j+\beta+2}}\left[(\cos\kappa -1)f_{0}\right] + \dfrac{(-1)^{j+1}}{(2j+\beta+1)\chi^{2j+\beta+1}}(\sin \kappa) f_{0}' \\
    &+ \dfrac{(-1)^{j}}{(2j+\beta)(2j+\beta+1)\chi^{2j+\beta}} \cos\kappa f_{0}''\Bigg\} \\
    &- F_{\phi;M} \dfrac{(-1)^{j}}{(2j+\beta)(2j+\beta+1)\chi^{2j+\beta}}f_{0} + \mathcal{O}\left(1/j^{3}\right).
    \end{split}
\end{align}
The $\sin\kappa$ terms vanish because $\cos\kappa - 1 = 0$ at leading order. This simplifies the equation for $f_{0}$ to:
\begin{equation}\label{eqn:f0_ODE}
    f_{0}'' - F_{\phi;M}(\phi_{0})f_{0} = 0.
\end{equation}
The two linearly independent solutions of \eqref{eqn:f0_ODE} are:
\begin{align*}
    f_{0}(\tilde{x}) &= \lambda_{k} g(\tilde{x}) = \lambda_{k}\phi_{0}'(\tilde{x}), \\
    f_{0}(\tilde{x}) &= \Lambda_{k}G(\tilde{x}) = \Lambda_{k} \phi_{0}'(\tilde{x}) \int_{\zeta}^{\tilde{x}}[\phi_{0}'(s)]^{-2} ds,
\end{align*}
where $\lambda_{k}$ and $\Lambda_{k}$ are constants that depend on $\kappa = \kappa_{k} = 2k\pi$.

The parameter $\beta$ can be determined by matching the pole orders in $\phi_{j}$ from \eqref{eqn:late_order_phi_n} and \eqref{eqn:late_order_phi_ansatz}. Since $\phi_{0}$ has a pole of order 1 at $\zeta$, it follows from \eqref{eqn:late_order_phi_n} that $\phi_{j}$ must have a pole of order $2j+1$ at $\zeta$. It is instructive to see the asymptotic behavior of $g(\tilde{x})$ and $G(\tilde{x})$ as $\tilde{x} \to \zeta$, they are
\begin{align} \label{eqn:G(x)_near_pole}
    g(\tilde{x}) \sim -\frac{\sqrt{2}}{(\tilde{x}-\zeta)^{2}}, \quad 
    G(\tilde{x}) \sim -\frac{1}{5\sqrt{2}}(\tilde{x}-\zeta)^{3}.
\end{align}
Matching the pole order of the ansatz for $f_{0} = \lambda_{k}g(\tilde{x})$ yields $\beta = -1$, while for $f_{0} = \Lambda_{k} G(\tilde{x})$ we find $\beta = 4$. 
Combining these contributions, the late-order terms $\phi_{j}$ are given by:
\begin{equation} \label{eqn:late_order_phi_j_expression}
    \phi_{j}(\tilde{x}) \sim \sum_{k=1}^{\infty}\left[\dfrac{(-1)^{j}\Gamma(2j-1)\lambda_{k}g(\tilde{x})}{[2k\pi(\tilde{x}-\zeta)]^{2j-1}} + \dfrac{(-1)^{j}\Gamma(2j+4)\Lambda_{k}G(\tilde{x})}{[2k\pi(\tilde{x}-\zeta)]^{2j+4}}\right].
\end{equation}

For real $\tilde{x}$, the dominant terms in $\phi_{j}$ come from $G(\tilde{x})$ and the nearest poles of $\phi_{0}$, namely $\zeta = \zeta_{0}$ and $\overline{\zeta}_{0}$. Thus, as $j \to \infty$, the dominant term in $\phi_{j}$ for real $\tilde{x}$ is:
\begin{align} \label{eqn:phij_dominant}
    \phi_{j}(\tilde{x}) &\sim \dfrac{(-1)^{j}\Gamma(2j+4)\Lambda_{1}G(\tilde{x})}{[2\pi(\tilde{x}-\zeta)]^{2j+4}}.
\end{align}
The constant $\Lambda_{1}$ can be determined by analyzing the behavior near the pole ($\tilde{x} \to \zeta$) as $j \to \infty$. To facilitate this, we introduce an inner region around the pole as follows.

Observe that as $\tilde{x} - \zeta = \mathcal{O}(\varepsilon)$, near $\zeta$, the leading order $\phi_{0}(\tilde{x})$ and the correction terms $\varepsilon^{2j}\phi_{j}(\tilde{x})$ are of order $\mathcal{O}(1/\varepsilon)$, thus making the expansion \eqref{eqn:expansion_phi} cease to be asymptotic as $\varepsilon \to 0$. To remedy this, we rescale the variables near $\zeta$ as
\begin{align}
    \tilde{x} = \zeta + \varepsilon q, \quad \varphi(q) = \varepsilon \phi(\tilde{x}).
\end{align}
Substituting these expressions into \eqref{eqn:dnls_stationary_real_eps} and retaining the leading-order terms as $\varepsilon \to 0$ (which are of $\mathcal{O}(\varepsilon^{-1})$), we obtain the inner equation:
\begin{align} \label{eqn:inner_eqn}
    \varphi(q+1) - 2\varphi(q) + \varphi(q-1) - \varphi^{3}(q) = 0.
\end{align}

We expand $\varphi(q)$ asymptotically as
\begin{align} \label{eqn:inner_expansion}
    \varphi(q) \sim \sum_{j = 0}^{\infty} \dfrac{A_{j}}{q^{2j+1}}, \quad \text{as} \quad |q| \to \infty.
\end{align}
This inner expansion must match the outer expansion \eqref{eqn:expansion_phi} as $\tilde{x} \to \zeta$. The leading-order matching condition gives $A_{0} = \sqrt{2}$. Substituting \eqref{eqn:inner_expansion} into \eqref{eqn:inner_eqn} and collecting terms at order $\mathcal{O}(q^{-(2j+3)})$, we obtain the recurrence relation:
\begin{equation} \label{eqn:recur_relation_Aj}
    0 = 2\sum_{p=1}^{j+1}\binom{2j+2}{2p}A_{j-p+1} - \sum_{k=0}^{j}\sum_{l=0}^{k}A_{l}A_{k-l}A_{j-k}.
\end{equation}

As $\tilde{x} \to \zeta$, using the asymptotic behavior of $G(\tilde{x})$ \eqref{eqn:G(x)_near_pole}, the dominant term in the outer expansion $\phi_{j}$ as $j \to \infty$ \eqref{eqn:phij_dominant} is asymptotically
\begin{align} \label{eqn:outer_expansion_near_zeta}
    \phi_{j} \sim \dfrac{(-1)^{j}\Gamma(2j+4)\Lambda_{1}G(\tilde{x})}{[2\pi(\tilde{x}-\zeta)]^{2j+4}} \sim \dfrac{-\sqrt{2}}{10}\dfrac{(-1)^{j}\Gamma(2j+4)\Lambda_{1}}{(2\pi)^{2j+4}(\tilde{x}-\zeta)^{2j+1}}.
\end{align} 
Matching the outer expansion \eqref{eqn:outer_expansion_near_zeta} with the inner expansion \eqref{eqn:inner_expansion} gives
\begin{align} \label{eqn:Lambda1}
    \Lambda_{1} = \dfrac{10}{\sqrt{2}}\lim_{j \to \infty} \dfrac{(-1)^{j+1}(2\pi)^{2j+4}}{\Gamma(2j+4)}A_{j}.
\end{align}
Numerical computation of the recurrence relation yields $\Lambda_{1} \approx -2533$.

\begin{figure}[tbhp]
    \centering
    \includegraphics[width=0.7\linewidth]{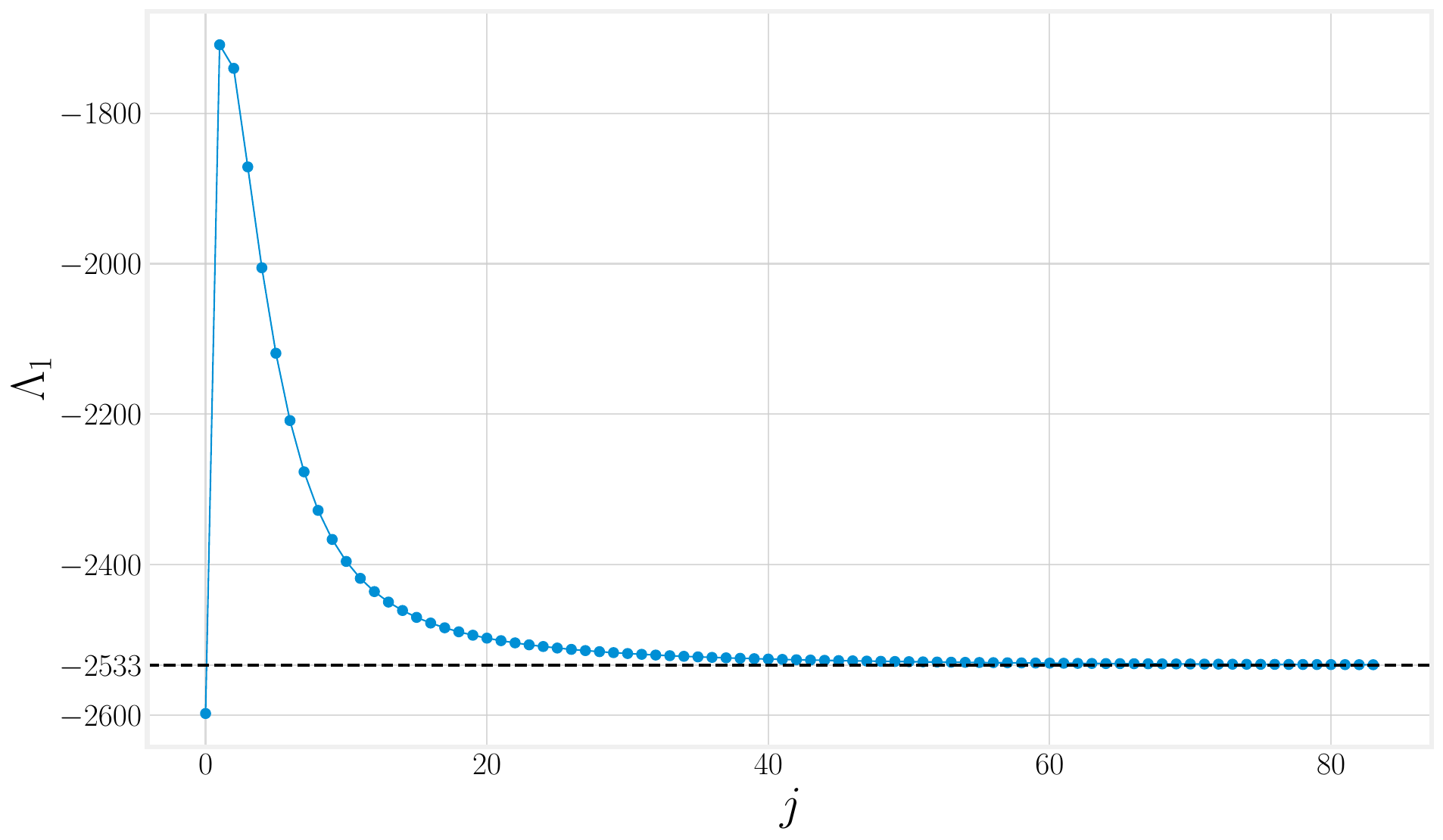}
    \caption{Approximate values of $\Lambda_{1}$ computed numerically.}
    \label{fig:L1_approx}
\end{figure}

\subsection{Optimal Truncation and Remainder Equation}
Having determined the dominant behavior of the diverging late order terms $\phi_{j}$ \eqref{eqn:phij_dominant}, it remains now to analyze the Stokes phenomenon, which triggers the activation of the exponentially small remainder term. This phenomenon is expected to occur along the Stokes lines \cite{dingle1973asymptotic}, which is a curve satisfying
\begin{equation}
    \text{Im}\ \chi = 0, \quad \text{Re}\ \chi \ge 0.
\end{equation} This gives the relevant Stokes lines as $\text{arg}(\tilde{x}-\zeta) = -\pi/2$ and $\text{arg}(\tilde{x}-\bar{\zeta}) = \pi/2$ from the poles $\zeta$ with $\text{Im}\ \zeta > 0$ and their conjugates. 
This requires truncation of the expansion \eqref{eqn:expansion_phi} at an optimal point, which can be deduced as follows. Suppose that the expansion \eqref{eqn:expansion_phi} is truncated after $N$ terms and denote the remainder of the truncated expansion by $R_{N}$, i.e. 
\begin{equation} \label{eqn:expansion_remainder}
    \phi = \sum_{j=0}^{N-1}\varepsilon^{2j}\phi_{j}(\tilde{x}) + R_{N}.
\end{equation}
The optimal truncation is the point $N \gg 1$ such that the remainder $R_{N}$ is minimized \cite{BerryHowls1990}. Following the heuristics of Boyd \cite{Boyd1999}, we choose $N$ such that : 
\begin{align*}
    \left|\dfrac{\varepsilon^{2N+2}\phi_{N+1}}{\varepsilon^{2N}\phi_{N}}\right| \sim 1,
\end{align*}
it is found that this condition leads to $N \sim \frac{|\kappa(\tilde{x}-\zeta)|}{2\varepsilon} + \nu$, with $\nu = O(1)$ ensuring $N \in \mathbb{Z}$.

Near the Maxwell point, we substitute \eqref{eqn:expansion_remainder} into \eqref{eqn:dnls_stationary_real_eps}, setting $\mu = \mu_{M} + \delta\mu$. The resulting equation for $R_{N}$ is
\begin{equation} \label{eqn:remainder_eqn}
    \Delta R_{N} - \varepsilon^{2}F_{\phi;M}(\phi_{0})R_{N} \sim \underbrace{\varepsilon^{2}\delta \mu F_{\mu;M}(\phi_{0})}_{\text{$\delta\mu$ forcing}} + \underbrace{2\sum_{m=1}^{\infty}\varepsilon^{2(N+m)}\sum_{p=m+1}^{N+m} \dfrac{1}{(2p)!}\partial_{\tilde{x}}^{2p}\phi_{N+m-p}}_{\text{Truncation forcing}}, \quad (\varepsilon \to 0).
\end{equation}

A WKB analysis of the homogeneous equation shows that \( R_{N} \) asymptotically takes the form \( R_{N} \sim e^{i\kappa_{k} \tilde{x}/\varepsilon}f_{0}(\tilde{x}) \) as \( \varepsilon \to 0 \). The dominant term in the late-order expansion of \( \phi_{j} \) \eqref{eqn:phij_dominant} suggests that this activated term will be proportional to \( G(\tilde{x}) \) with \( \kappa_{k} = 2\pi \), across the corresponding Stokes line. We will analyze the contribution from each of the forcing terms separately. We will refer to the forcing due to the presence of $\delta\mu$ as $\delta\mu$ forcing and the forcing term due to the truncation of the asymptotic series \eqref{eqn:expansion_phi} as the truncation forcing term. The contribution from each of the forcing term to the remainder will be denoted respectively as $R_{N,\delta\mu}$ and $R_{N,\text{trunc}}$.

To analyze the effect of $\delta\mu$ forcing, we let $R_{N} = \delta\mu P(\tilde{x})$. Substituting this ansatz into \eqref{eqn:remainder_eqn} gives the leading-order equation for $P(\tilde{x})$ as
\begin{equation}
    P'' - F_{\phi;M}(\phi_{0})P = F_{\mu;M}(\phi_{0}),
\end{equation}
which can be solved to yield
\begin{equation}\label{eqn:Ptildex}
    P(\tilde{x}) = \phi_{0}'(\tilde{x}) \int^{\tilde{x}}[\phi_{0}'(t)]^{-2} \left[\int_{0}^{\phi_{0}(t)}F_{\mu;M}(\xi)d\xi\right] dt.
\end{equation}
It is then found that $R_{N,\delta\mu}$ is given by
\begin{equation}\label{eqn:Rn_delta_forcing}
    R_{N,\delta}(\tilde{x}) = \delta\mu P(\tilde{x}) = \delta\mu \phi_{0}'(\tilde{x}) \int^{\tilde{x}}[\phi_{0}'(s)]^{-2} \left[\int_{0}^{\phi_{0}(t)}F_{\mu;M}(\xi)d\xi\right] ds.
\end{equation}

We will now focus on the contribution from the truncation forcing term.
Using the general late-order form of $\phi_{j}$ \eqref{eqn:late_order_phi_ansatz}, Stirling's approximation for the Gamma function : 
\[
\Gamma(z) \sim \sqrt{2\pi}z^{z-\frac{1}{2}}e^{-z} \ \ (z \gg 1),
\]
and after reindexing the sum in the truncation forcing term 
to start at $m = 0$, we have: 
\begin{align*}
    &2\sum_{m=1}^{\infty}\varepsilon^{2(N+m)}\sum_{p=m+1}^{N+m} \dfrac{1}{(2p)!}\partial_{\tilde{x}}^{2p}\phi_{N+m-p} \\
    \sim \ &2 \sum_{m=0}^{\infty}\varepsilon^{2N+2}\varepsilon^{2m} \sum_{p=m+2}^{N+m+1}\dfrac{(-1)^{N+m+1+p}}{(2p)!}\dfrac{\sqrt{2\pi}(2N+\beta+2m+2)^{2N+\beta+2m+3/2}e^{-(2N+\beta+2m+2)}}{\chi^{2N+\beta+2m+2}}.
\end{align*}
Since $2N \gg \beta + 2m + 2$, we take the following approximation
\[
(2N+\beta+2m+2)^{2N+\beta+2m+2-1/2}e^{-(2N+\beta+2m+2)} \sim (2N)^{2N+\beta+2m+2-1/2}e^{-2N},
\]
and pull this factor out of the summation. After reversing the order of summation, we can compute the sum exactly. The truncation forcing term then becomes
\begin{align*}
    &\sim \dfrac{-2\sqrt{2\pi}(-1)^{N+1}\varepsilon^{2N+2}(2N)^{2N+\beta+3/2}e^{-2N}}{\chi^{2N+\beta+2}}\left[\sum_{p=2}^{\infty}\dfrac{(-1)^{p}\kappa^{2p}}{(2p)!}\sum_{m=0}^{p-2}(-1)^{m}\left(\dfrac{\varepsilon(2N)}{\chi}\right)^{2m}\right]f_{0} \\
    &\sim \dfrac{-2\sqrt{2\pi}(-1)^{N}\varepsilon^{2N}(2N)^{2N+\beta-1/2}e^{-2N}}{\chi^{2N+\beta+2}[\chi^{2}+(2N\varepsilon)^{2}]}\left(\cosh\left(\dfrac{2\kappa N\varepsilon}{\chi}\right) - 1\right).
\end{align*}
As we expect a Stokes phenomenon to occur as we cross the Stokes lines passing through the pole $\zeta$, we also anticipate an abrupt change in the remainder term $R_{N,\text{trunc}}$ around the neighborhood of $\zeta$. To this end, following the ideas of Olde Daalhuis \textit{et al}.\ \cite{daalhuis1995}, we introduce the polar coordinates 
and write $\chi = \kappa(\tilde{x}-\zeta) = \rho e^{i\theta}$. This is done such that we may observe the variation in $R_{N,\text{trunc}}$ with respect to $\theta$, which happens much more rapidly than the variation in $\rho$. Under the polar coordinates, $N$ becomes $N = \frac{\rho}{2\varepsilon} + \nu$. The prefactor in the truncation forcing term then becomes
\begin{align*}
    \dfrac{\varepsilon^{2N}(2N)^{2N+\beta-1/2}e^{-2N}}{[\kappa(\tilde{x}-\zeta)]^{2N+\beta+2}[(\kappa(\tilde{x}-\zeta))^{2}+(2N\varepsilon)^{2}]} 
    &\sim \dfrac{\varepsilon^{2N}(2N)^{2N+\beta-1/2}e^{-2N}}{\rho^{2N+\beta-2}e^{i(2N+\beta-2)\theta}[\rho^{2}e^{i2\theta}+4\varepsilon^{2}N^{2}]} \\
    &\sim \dfrac{\varepsilon^{1/2-\beta}e^{-i\theta(2N-2+\beta)}e^{-\rho/\varepsilon}}{\rho^{1/2}(e^{i2\theta} + 1)} \quad \text{(as $N \to \infty$)}.
\end{align*}
The truncation forcing term is then
\begin{align*}
    \sim (-1)^{N}2\sqrt{2\pi}\dfrac{\varepsilon^{1/2-\beta}e^{-i\theta(2N-2+\beta)}e^{-\rho/\varepsilon}}{\rho^{1/2}}\dfrac{\cosh(\kappa e^{-i\theta}) - 1}{e^{i2\theta} + 1}f_{0}.
\end{align*}
To solve for the remainder due to the truncation forcing term, we set $R_{N,\text{trunc}}(\tilde{x}) = e^{\pm i\kappa (\tilde{x}-\zeta)/\varepsilon}S_{N}(\tilde{x})$ with $\kappa = 2k\pi$. Substituting this into \eqref{eqn:remainder_eqn}, the left-hand side becomes
\begin{align*}
    &e^{\pm i\kappa (\tilde{x}-\zeta)/\varepsilon}\left\{2(\cos\kappa - 1)S_{N} \pm 2i \sin \kappa \varepsilon S_{N}' + \varepsilon^{2}[\cos \kappa S_{N}'' - F_{\phi;M}(\phi_{0})S_{N}] + \mathcal{O}(\varepsilon^{3})\right\} \\
    &\sim e^{\pm i\kappa (\tilde{x}-\zeta)/\varepsilon}\left[\varepsilon^{2}(S_{N}''-F_{\phi;M}(\phi_{0})S_{N}) + \mathcal{O}(\varepsilon^{3}S_{N})\right].
\end{align*}
After moving the exponential factor to the right-hand side, the equation for $S_{N}(\tilde{x})$ at leading order becomes
\begin{align} \label{eqn:SN_eqn}
    S_{N}''-F_{\phi;M}(\phi_{0})S_{N} \sim \dfrac{2\sqrt{2\pi}(-1)^{N}e^{-i\theta(2N-2+\beta)}}{\varepsilon^{\beta+\frac{3}{2}} \rho^{1/2}} \dfrac{\cosh(\kappa e^{-i\theta}) - 1}{e^{i2\theta} + 1} f_{0} \exp[\mp i \rho e^{i\theta}/\varepsilon - \rho/\varepsilon].
\end{align}
From the truncation condition, the terms proportional to $\exp\left[\mp i\rho e^{i\theta}/\varepsilon - \rho/\varepsilon\right]$ dominate except when $\theta = \pm \pi/2$, where they become algebraic in $\varepsilon$. This signifies the occurrence of the Stokes phenomenon when the Stokes lines associated with $\theta = \pm \pi/2$ are crossed. For poles $\zeta$ with $\text{Im }\zeta > 0$, the Stokes line corresponds to $\theta = -\pi/2$, while for poles with $\text{Im }\zeta < 0$, it corresponds to $\theta = \pi/2$. By symmetry, the contribution of $\bar{\zeta}$ is the complex conjugate of that from $\zeta$, so we focus on $\zeta$ with $\text{Im }\zeta > 0$. The truncated remainder term is then
\begin{equation}
    R_{N,\text{trunc}} = e^{-i\kappa (\tilde{x}-\zeta)/\varepsilon}S_{N}(\tilde{x}).
\end{equation}

To analyze the manifestation of the Stokes phenomenon near the Stokes line, we define an inner angular variable $\tilde{\theta}$ and set $\theta = -\pi/2 + \eta(\varepsilon)\tilde{\theta}$. This substitution gives the equation for $S_{N}$ as
\begin{align}
    S_{N}'' - F_{\phi;M}(\phi_{0})S_{N} \sim \dfrac{2\sqrt{2\pi}}{\varepsilon^{\beta+\frac{3}{2}}\rho^{1/2}}(-1)^{N}e^{-i\theta(2N-2+\beta)}e^{i\rho e^{i\theta}/\varepsilon}e^{-\rho/\varepsilon} h(\theta) f_{0},
\end{align}
where
\begin{align}
    h(\theta) = \dfrac{\cosh(\kappa e^{-i\theta}) - 1}{e^{i2\theta} + 1}.
\end{align}

Considering the prefactor $(-1)^{N}{e^{-i\theta(2N-2+\beta)}\rho^{-1/2}}e^{i\rho e^{i\theta}/\varepsilon}e^{-\rho/\varepsilon}$, we have
\begin{align*}
    (-1)^{N}e^{i\theta(2N-2+\beta)}e^{i\rho e^{i\theta}/\varepsilon}e^{-\rho/\varepsilon}
    &= \exp\Bigg[-i\pi N + i\theta(2N-2+\beta)+ie^{i\theta}\frac{\rho}{\varepsilon} - \frac{\rho}{\varepsilon}\Bigg] \\
    &= \exp\Bigg[-i\pi N + i\left(-\frac{\pi}{2}+\eta \tilde{\theta}\right)\left(2N - 2 + \beta\right) +i e^{i(-\pi/2 + \eta\tilde{\theta})}\frac{\rho}{\varepsilon} - \frac{\rho}{\varepsilon}\Bigg] \\
    &= \exp\Bigg[i\pi-i\frac{\pi}{2}\beta + i \eta\tilde{\theta}2N + i\eta\tilde{\theta}(\beta-2) + \frac{\rho}{\varepsilon}\left(e^{i\eta\tilde{\theta}}- 1\right)\Bigg] \\
    &\sim \exp\Bigg[i\pi-i\frac{\pi}{2}\beta + i\eta\tilde{\theta}\left(\frac{\rho}{\varepsilon} + 2\nu\right) + i\eta\tilde{\theta}(\beta-2) + \frac{\rho}{\varepsilon}\left(i\eta\tilde{\theta} - \frac{\eta^{2}\tilde{\theta}^{2}}{2} + O(\eta^{3})\right)\Bigg] \\
    &\sim \exp\Bigg[i\pi - i\frac{\pi}{2}\beta - \frac{\eta^{2}\tilde{\theta}^{2}}{2}\frac{\rho}{\varepsilon} + i\eta\tilde{\theta}(2\nu+\beta-2) + O\left(\frac{\eta^{3}}{\varepsilon}\right)\Bigg].
\end{align*}

Balancing terms in the exponent are achieved when \(\eta = \sqrt{\varepsilon}\). With this scaling, we can neglect higher-order terms, leaving
\[
(-1)^{N}e^{i\theta(2N-2+\beta)}e^{i\rho e^{i\theta}/\varepsilon}e^{-\rho/\varepsilon} \sim -e^{i\beta\pi/2} e^{-\rho\tilde{\theta}^{2}/2}.
\]
As \(\varepsilon \to 0\), the function \(h(\theta)\) asymptotes to:
\begin{align*}
    \frac{\cosh(\kappa e^{-i\theta}) - 1}{e^{i2\theta} + 1} &= \frac{\cosh(i\kappa e^{-i\sqrt{\varepsilon}\tilde{\theta}}) - 1}{1 - e^{i2\sqrt{\varepsilon}\tilde{\theta}}} \\
    &= \frac{\cos(\kappa e^{-i\sqrt{\varepsilon}\tilde{\theta}}) - 1}{1 - e^{i2\sqrt{\varepsilon}\tilde{\theta}}} \\
    &\sim \frac{i\kappa^{2}}{4}\sqrt{\varepsilon}\tilde{\theta}.
\end{align*}
The leading-order equation for \(S_N\) then becomes:
\begin{align}\label{eqn:SN_eqn_inner_theta}
    S_N'' - F_{\phi;M}(\phi_0) S_N \sim -i \sqrt{\frac{\pi}{2}} \rho^{-1/2}
    e^{i \beta \pi / 2} \varepsilon^{-1-\beta} \kappa^2 \tilde{\theta} e^{-\rho \tilde{\theta}^2 / 2} f_0.
\end{align}
In terms of \(\tilde{\theta}\), the derivative transforms as:
\[
\partial_{\tilde{x}} = \frac{\kappa}{\rho} e^{-i\sqrt{\varepsilon}\tilde{\theta}} \varepsilon^{-1/2} \partial_{\tilde{\theta}},
\]
so that:
\[
S_N'' \sim \frac{\kappa^2}{\rho^2} \varepsilon^{-1} S_{N,\tilde{\theta}\tilde{\theta}}.
\]
Substituting this expression, we find:
\[
S_{N,\tilde{\theta}\tilde{\theta}} \sim -i \sqrt{\frac{\pi}{2}} \rho^{3/2} e^{i\beta\pi/2} \varepsilon^{-\beta} \tilde{\theta} e^{-\rho\tilde{\theta}^2/2} f_0.
\]
If we write \(S_N = \varepsilon^{-\beta} f_0(\tilde{x}) \tilde{S}_N(\tilde{\theta})\) and note that \(f_0\) varies slowly with \(\tilde{\theta}\), then:
\[
\tilde{S}_{N,\tilde{\theta}\tilde{\theta}} \sim -i \sqrt{\frac{\pi}{2}} e^{i\beta\pi/2} \rho^{3/2} \tilde{\theta} e^{-\rho\tilde{\theta}^2/2}.
\]
After integrating twice and applying the boundary condition \(\tilde{S}_N \to 0\) as \(\tilde{\theta} \to -\infty\), we have:
\begin{align} \label{eqn:SN_tilde}
    \tilde{S}_N(\tilde{\theta}) \sim \frac{i\pi}{2} e^{i\beta\pi/2} \text{erfc}\left(-\tilde{\theta}\sqrt{\frac{\rho}{2}}\right),
\end{align}
where \(\text{erfc}(\tilde{x})\) is the complementary error function:
\[
\text{erfc}(\tilde{x}) = \frac{2}{\sqrt{\pi}} \int_{\tilde{x}}^{\infty} e^{-y^2} \, dy.
\]
Crossing the Stokes line from \(\text{Re}(\tilde{x}) < 0\) to \(\text{Re}(\tilde{x}) > 0\) results in a jump:
\[
\tilde{S}_N(\theta \to (-\pi/2)^+) - \tilde{S}_N(\theta \to (-\pi/2)^-) \sim i\pi e^{i\beta\pi/2}.
\]
This jump corresponds to the pole \(\zeta\). From \eqref{eqn:SN_tilde}, we observe that the Stokes phenomenon occurs smoothly, though rapidly, as \(\text{arg}(\tilde{x}-\zeta) - (-\pi/2) = O(\sqrt{\varepsilon})\).

Since the dominant contribution to the late-order terms is \(f_0 = \Lambda_1 G(\tilde{x})\), the pole \(\zeta\) with \(\text{Im}\zeta > 0\) contributes to the remainder as:
\begin{align} \label{eqn:RN_trunc}
    R_{N,\text{trunc}} \sim \frac{i\pi}{2} e^{i2\pi\left(n_0 + \frac{\zeta}{\varepsilon}\right)} \varepsilon^{-4} \text{erfc}\left(-\tilde{\theta}\sqrt{\frac{\rho}{2}}\right) \Lambda_1 G(\tilde{x}),
\end{align}
while the contribution from \(\overline{\zeta}\) is the complex conjugate of \eqref{eqn:RN_trunc}.

The dominant remainder contribution arises from the poles \(\zeta = \zeta_0 = i\frac{3\pi}{\sqrt{2}}\) and its conjugate \(\overline{\zeta}_0\), which are nearest to the real axis. Combining \(R_{N,\delta}\) and \(R_{N,\text{trunc}}\), the total remainder is:
\begin{align} \label{eqn:RN}
    R_N \sim \delta \mu P(\tilde{x}) + \frac{i\pi}{2} e^{i2\pi n_0} \varepsilon^{-4} e^{-3\sqrt{2}\pi^2/\varepsilon} \text{erfc}\left(-\tilde{\theta}\sqrt{\frac{\rho}{2}}\right) \Lambda_1 G(\tilde{x}) + \text{c.c.}
\end{align}
For \(\tilde{x} > 0\), this simplifies to:
\begin{align} \label{eqn:RN_x>0}
    R_N \sim \delta \mu P(\tilde{x}) + i\pi \varepsilon^{-4} e^{-3\sqrt{2}\pi^2/\varepsilon} e^{i2\pi n_0} \Lambda_1 G(\tilde{x}) + \text{c.c.}
\end{align}

\subsection{Pinning Width Approximation}
As we have seen, the leading contributions to the remainder \(R_N\) as the Stokes line is crossed are given by the functions \(G(\tilde{x})\) and \(P(\tilde{x})\). For \(\tilde{x} \gg 1\), these terms grow exponentially as 
\[
G(\tilde{x}) = \phi_0'(\tilde{x}) \int_{\zeta}^{\tilde{x}} [\phi_0'(s)]^{-2} \, ds \sim \frac{27}{8} e^{\sqrt{2}\tilde{x}/3},
\]
\[
P(\tilde{x}) = \phi_0'(\tilde{x}) \int^{\tilde{x}} [\phi_0'(s)]^{-2} \left[\int_{0}^{\phi_0(s)} F_{\mu;M}(\xi) \, d\xi\right] ds \sim -\frac{3}{4} e^{\sqrt{2}\tilde{x}/3}.
\]
Thus, the dominant terms in the remainder as \(\tilde{x} \to \infty\) are:
\begin{align*}
    R_N &\sim \delta \mu P(\tilde{x}) - i\pi \varepsilon^{-4} e^{-3\sqrt{2}\pi^2/\varepsilon} |\Lambda_1| e^{i2\pi n_0} G + \text{c.c.} \\
    &\sim \delta \mu P(\tilde{x}) + 2\pi \varepsilon^{-4} e^{-3\sqrt{2}\pi^2/\varepsilon} |\Lambda_1| \left[\sin(2\pi n_0) \text{Re } G(\tilde{x}) + \cos(2\pi n_0) \text{Im } G(\tilde{x})\right] \\
    &\sim \left[-\frac{3}{4} \delta \mu + \frac{27}{4} \pi \varepsilon^{-4} e^{-3\sqrt{2}\pi^2/\varepsilon} |\Lambda_1| \sin(2\pi n_0)\right] e^{\sqrt{2}\tilde{x}/3}.
\end{align*}
The remainder term is bounded as \(\tilde{x} \to \infty\) if the exponentially growing component vanishes. This condition is satisfied if:
\begin{equation} \label{eqn:dmu_onefront}
    \delta \mu = 9\pi \varepsilon^{-4} e^{-3\sqrt{2}\pi^2/\varepsilon} |\Lambda_1| \sin(2\pi n_0).
\end{equation}
Therefore, stationary front solutions exist if:
\begin{equation} \label{eqn:pinning_region_delta_mu}
    |\delta \mu| \leq 9\pi \varepsilon^{-4} e^{-3\sqrt{2}\pi^2/\varepsilon} |\Lambda_1|.
\end{equation}
In this range of \(\delta \mu\), the stationary fronts of \eqref{eqn:dnls_main} remain pinned to the lattice, defining the pinning region for the stationary front.

\begin{figure}[tbhp]
    \centering
    \includegraphics[width=0.6\linewidth]{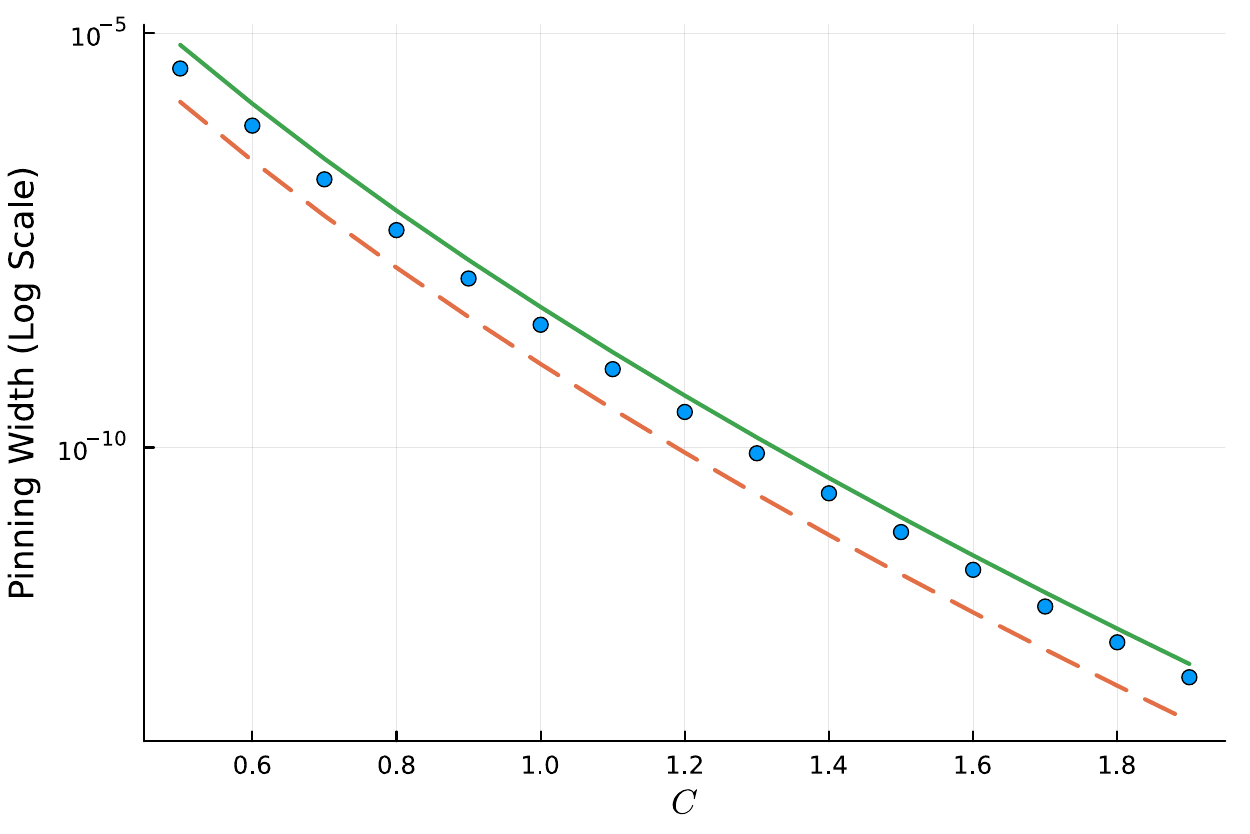}
    \caption{Comparison between the numerically obtained pinning width and the
asymptotic approximations for \( C = 0.5 \) to \( C = 1.9 \). The blue dots represent the numerical results. The green solid line corresponds
to the approximation derived from \eqref{eqn:pinning_region_delta_mu},
while the orange dashed line illustrates the variational approximation
obtained in \cite{kusdiantara2024analysis}.
}
    \label{fig:pinning_width}
\end{figure}

The asymptotic approximation of the pinning width, denoted as
\(W = 2\max |\delta \mu|\), is compared with numerical results in Figure~\ref{fig:pinning_width}. Rewriting this approximation in terms of the original variable \(C\) rather than \(\varepsilon\), we find:
\begin{equation} \label{eqn:pinning_width_approx_exp_asymp}
    W = 2\max |\delta \mu| \sim \dfrac{9\pi}{2}|\Lambda_1| C^{2}e^{-3\pi^{2}\sqrt{C}} \approx 11398.5 C^{2}e^{-3\pi^{2}\sqrt{C}}.
\end{equation}
The numerical pinning width is determined using numerical continuation of the one-front solution as \(\mu\) varies for different values of \(C\). We also compare this approximation to a variational approximation from \cite{kusdiantara2024analysis}. Both approaches have the same order of magnitude, \(\mathcal{O}(C^{2}e^{-3\pi^{2}\sqrt{C}})\), but the prefactor in \cite{kusdiantara2024analysis} is \(24\pi^{5} \approx 7344.47\). The approximation obtained from exponential asymptotics, however, aligns more closely with the numerical results than the variational approximation.

\subsection{Two-Front Solutions}
Having established the necessary conditions for the existence of
stationary one-front solutions, we now construct localized stationary
solutions in the form of a droplet. This soliton consists of two
matched one-front solutions: an up-front solution \(\phi(\varepsilon(n-n_{0}))\)
(as derived above) is paired with a down-front solution
\(\phi(-\varepsilon(n-n_{0}) + L/\varepsilon)\) in the far-field.
The down-front solution is obtained by rotating the up-front solution and
shifting it by a factor of \(L/\varepsilon\), where \(L = \mathcal{O}(1)\) is a
positive constant. This effectively rotates and shifts the center of the front
(\(n_{0}\)) to \(-n_{0} - L/\varepsilon^{2}\).
The shift scale \(L/\varepsilon\) is chosen because the remainder term
fails to be exponentially small when \(\tilde{x} = \mathcal{O}(1/\varepsilon)\)
and \(\tilde{x} > 0\). In practice, we match the up-front solution
\(\phi(\tilde{x})\) with the down-front solution
\(\phi(-\tilde{x} + L/\varepsilon)\) in the far-field matching region,
which corresponds to \(\tilde{x} \gg 1\) and \(-\tilde{x} + L/\varepsilon \gg 1\).

In the far-field (\(\tilde{x} \gg 1\)), the up-front solution takes the form:
\[
\phi \sim \frac{2}{3}\left(1 - e^{-\sqrt{2}\tilde{x}/3}\right) + 
\left[\frac{27}{4}\pi \varepsilon^{-4}e^{-3\sqrt{2}\pi^{2}/\varepsilon}|\Lambda_{1}|\sin(2\pi n_{0}) - \frac{3}{4}\delta\mu\right]e^{\sqrt{2}\tilde{x}/3}.
\]
In the far-field of the down-front solution (\(-\tilde{x} + L/\varepsilon \gg 1\)), we have:
\[
\phi \sim \frac{2}{3}\left(1 - e^{-\sqrt{2}(-\tilde{x}+L/\varepsilon)/3}\right) + 
\left[\frac{27}{4}\pi \varepsilon^{-4}e^{-3\sqrt{2}\pi^{2}/\varepsilon}|\Lambda_{1}|\sin\left(2\pi \left(-n_{0}-\frac{L}{\varepsilon^{2}}\right)\right) - \frac{3}{4}\delta\mu\right]e^{\sqrt{2}(-\tilde{x}+L/\varepsilon)/3}.
\]

Matching the terms proportional to \(e^{-\sqrt{2}\tilde{x}/3}\) from the up-front and the down-front solutions yields:
\begin{equation} \label{eqn:matching_e-alpha}
    -\frac{2}{3}e^{-\sqrt{2}L/3\varepsilon} = 
    \frac{27}{4}\pi\varepsilon^{-4}e^{-3\sqrt{2}\pi^{2}/\varepsilon}|\Lambda_{1}|\sin\left(2\pi \left(-n_{0}-\frac{L}{\varepsilon^{2}}\right)\right) - \frac{3}{4}\delta\mu.
\end{equation}

Similarly, the terms proportional to \(e^{\sqrt{2}\tilde{x}/3}\) gives:
\begin{equation} \label{eqn:matching_e+alpha}
    -\frac{2}{3}e^{-\sqrt{2}L/3\varepsilon} = 
    \frac{27}{4}\pi \varepsilon^{-4}e^{-3\sqrt{2}\pi^{2}/\varepsilon}|\Lambda_{1}|\sin(2\pi n_{0}) - \frac{3}{4}\delta\mu.
\end{equation}

Equating these expressions shows that the sine terms must be equal, i.e. 
\[
\sin(2\pi n_{0}) = \sin\left(2\pi \left(-n_{0}-\frac{L}{\varepsilon^{2}}\right)\right).
\]

This condition is satisfied if either:
\begin{equation} \label{eqn:n0_condition_snake}
    n_{0} = -\frac{L}{2\varepsilon^{2}} + \frac{k}{2},
\end{equation}
or:
\begin{equation} \label{eqn:L/eps_condition_ladder}
    \frac{L}{\varepsilon} = \varepsilon\left(k - \frac{1}{2}\right),
\end{equation}
where \(k\) is an integer of order \(\mathcal{O}(\varepsilon^{-2})\) and \(L = \mathcal{O}(1)\).

Substituting either \eqref{eqn:n0_condition_snake} or \eqref{eqn:L/eps_condition_ladder} into \eqref{eqn:matching_e+alpha} provides an implicit relationship between the other unknown (\(L/\varepsilon\) in the case of \eqref{eqn:n0_condition_snake} and \(n_{0}\) in the case of \eqref{eqn:L/eps_condition_ladder}) and \(\delta\mu\):
\begin{equation} \label{eqn:dmu_snake_asymp}
\delta\mu = \frac{8}{9}e^{-\sqrt{2}L/3\varepsilon} + 9\pi \varepsilon^{-4} e^{-3\sqrt{2}\pi^{2}/\varepsilon}|\Lambda_{1}|\sin\left[\pi\left(\frac{-L}{\varepsilon^{2}} + k\right)\right],
\end{equation}
or:
\begin{equation} \label{eqn:dmu_ladder_asymp}
\delta\mu = \frac{8}{9}e^{-\sqrt{2}L/3\varepsilon} + 9\pi \varepsilon^{-4} e^{-3\sqrt{2}\pi^{2}/\varepsilon}|\Lambda_{1}|\sin(2\pi n_{0}).
\end{equation}

As \(L \to \infty\), the pinning region for \(\delta\mu\) converges to that of the one-front solution, meaning that the pinning region for one-front and double-front solutions is identical in this limit.

The solution types depend on the parity of \(k\) in \eqref{eqn:n0_condition_snake}. If \(k\) is even, the center of the localized solution (\(n = n_{0} + L/2\varepsilon^{2}\)) lies onsite, whereas if \(k\) is odd, it is intersite (located between lattice sites). These solutions, parameterized by \eqref{eqn:n0_condition_snake}, are known as "snakes." The solutions parameterized by \eqref{eqn:L/eps_condition_ladder} are called "ladders," as their bifurcation structure resembles ladder rungs between the snakes.

A composite solution can then be constructed by summing the up-front and down-front solutions while subtracting their overlapping parts in the matching region:
\begin{equation} \label{eqn:phi_two_front_asymptotic}
    \phi(\tilde{x}) \sim \phi_{0}(\tilde{x}) + R_{N}(\tilde{x};n_{0}) + \phi_{0}(-\tilde{x}+L/\varepsilon) + R_{N}\left(-\tilde{x}+\frac{L}{\varepsilon}; -n_{0}-\frac{L}{\varepsilon^{2}}\right) - \phi_{\text{overlap}}(\tilde{x}),    
\end{equation}
where \(R_{N}(\tilde{x};n_{0})\) captures the dominant far-field contributions from \(G(\tilde{x})\) and \(P(\tilde{x})\), and
\[
\phi_{\text{overlap}}(\tilde{x}) = \frac{2}{3}\left(1 - e^{-\sqrt{2}\tilde{x}/3} - e^{-\sqrt{2}L/3\varepsilon}e^{\sqrt{2}\tilde{x}/3}\right).
\]
The formulae for the snakes \eqref{eqn:dmu_snake_asymp} and ladders \eqref{eqn:dmu_ladder_asymp}, along with the asymptotic composite solution \eqref{eqn:phi_two_front_asymptotic} are used to obtained the approximate snakes and ladders in Fig. \ref{fig:bifurcation_diag_C=0.5}. To obtain this plot, we manually eliminate the growing terms in \eqref{eqn:phi_two_front_asymptotic} according to \eqref{eqn:dmu_snake_asymp} and \eqref{eqn:dmu_ladder_asymp}. Otherwise, numerical errors would leave residual factors which would cause the asymptotic solution to blow-up. To mitigate these erroneous blow-ups, we consider $\phi_{0}$ as 
\[
\phi_{0}(\tilde{x}) = \frac{1}{3}\left[1 + \tanh\left(\frac{x}{3\sqrt{2}}\right)\right]
\]
and the composite solution is computed as
\begin{align*}
    \left[\phi_{0}(\tilde{x}) + \phi_{0}(-\tilde{x} + L/\varepsilon) - \frac{2}{3}\right] + \left[R_{N}(\tilde{x};n_{0}) - \frac{-2}{3}e^{\sqrt{2}(\tilde{x} - L/\varepsilon)/3}\right] + \left[R_{N}\left(-\tilde{x}+\frac{L}{\varepsilon}; -n_{0}-\frac{L}{\varepsilon^{2}}\right) - \frac{-2}{3}e^{-\sqrt{2}\tilde{x}/3}\right],
\end{align*}
where each term in the brackets are computed separately and then added together.

By modifying \((n-n_{0}) \mapsto -(n-n_{0})\), the same approach can be used to construct bubble soliton solutions. The resulting pinning region for bubbles is identical to that for droplets.

\section{Linear Stability}\label{sec4}

 \begin{figure}[tbhp]
     \centering
     \includegraphics[width=1\linewidth]{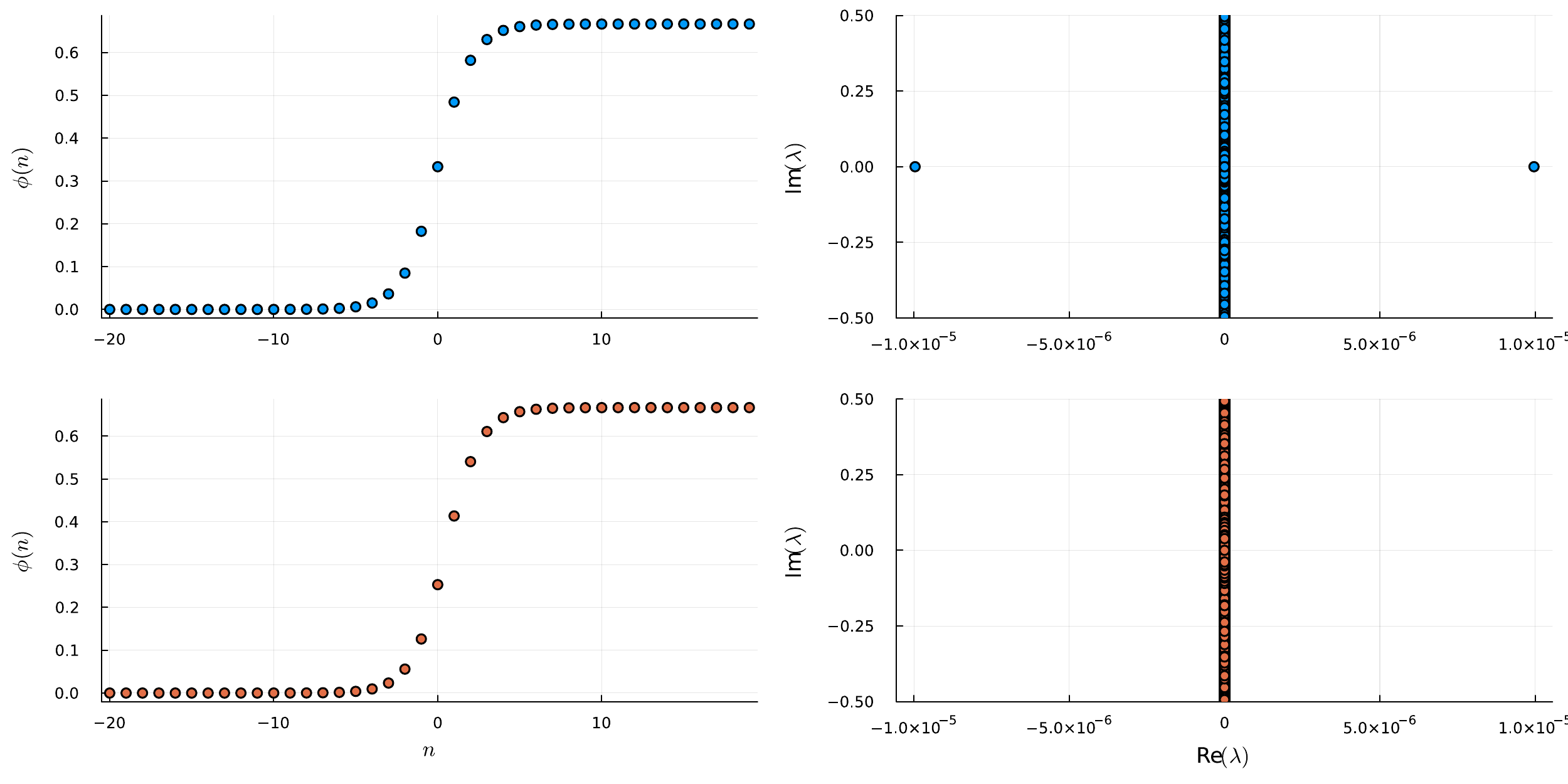}
     \caption{Examples of Maxwell fronts and their associated spectra for \( C = 0.5 \) are shown. The top panels display the profile and spectrum of the unstable onsite Maxwell front, while the bottom panels present the corresponding plots for the stable intersite Maxwell front. }
     \label{fig:kink_on_off_spectrum}
 \end{figure}

As the two types of droplets, onsite and intersite, move up along the snaking bifurcation diagram, they interchange stability. Up in the bifurcation diagram, a pair of critical eigenvalues determine the stability of the droplets and bubbles \cite{kusdiantara2024analysis}. The difference between these eigenvalues is exponentially small in the separation distance between the fronts. Consequently, the stability of the droplets and bubbles is primarily governed by the stability of the fronts that form them. Therefore, in the following analysis, we will focus on studying the stability of the one-front solution, particularly at the Maxwell point ($\mu = \mu_{M} = -2/9$). We will refer to this solution as the Maxwell front. There are also two types of Maxwell fronts: onsite and intersite. We will show that the intersite front is stable, while the onsite front is unstable; see Fig.\ \ref{fig:kink_on_off_spectrum}. This result is particularly interesting because it is different from the common dark solitons in the cubic discrete NLS equation where the stability is the other way around \cite{susanto2005discrete,fitrakis2007dark,pelinovsky2008stability,maluckov2007dark}. 

Linearizing the discrete NLS equation \eqref{eqn:dnls_main} about the real stationary solution $\tilde{\phi}$, we introduce a perturbation of the form $\phi = \tilde{\phi} + (p + iq)$, where $p$ and $q$ are small perturbations satisfying $|p|, |q| \ll 1$. Setting $p = u e^{\lambda t}$ and $q = v e^{\lambda t}$, we obtain the linear eigenvalue problem
\begin{align} \label{eqn:eigval_eqn}
    \lambda \begin{bmatrix}
        u \\ v
    \end{bmatrix} = \begin{bmatrix}
        O & L_{-} \\ -L_{+} & O
    \end{bmatrix}\begin{bmatrix}
        u \\ v
    \end{bmatrix} = N \begin{bmatrix}
        u \\ v
    \end{bmatrix},
\end{align}
where
\begin{align}
    L_{-} &= -\dfrac{1}{\varepsilon^{2}}\Delta - \mu - |\tilde{\phi}| + |\tilde{\phi}|^{2}, \label{eqn:Lm} \\
    L_{+} &= -\dfrac{1}{\varepsilon^{2}}\Delta + F_{\phi}(\tilde{\phi};\mu). \label{eqn:Lp}
\end{align}
This system can be rewritten as
\begin{equation} \label{eqn:eigval_eqn_LmLp}
    L_{+}L_{-}v = -\lambda^{2} v.
\end{equation}

We will consider linearizing around the Maxwell front, denoted as $\tilde{\phi}_{M}$. As $\varepsilon \to 0$, it is asymptotically given by
\begin{equation}\label{eqn:phi_M}
    \tilde{\phi}_{M}(\tilde{x}) \sim \phi_{0}(\tilde{x}) + 2\pi \varepsilon^{-4}e^{-3\sqrt{2}\pi^{2}/\varepsilon}|\Lambda_{1}|\left[\text{Re }G(\tilde{x})\sin(2\pi n_{0}) + \text{Im }G(\tilde{x}) \cos(2\pi n_{0})\right]
\end{equation}
as $\varepsilon \to 0$, where $n_{0} = 0$ (onsite front) or $n_{0} = 1/2$ (intersite front). In determining the stability of this solution, we will make use of the following lemma regarding the eigenvalues of the operator $L_{+}$.

\begin{lemma} \label{lem:eigval_Lp}
    Let $\tilde{\phi}_{M}$ be the Maxwell front to \eqref{eqn:dnls_main} and let $\varepsilon \to 0$. The corresponding operator $L_{+}$ has exactly one negative eigenvalue in the case of an onsite Maxwell front and exactly one positive eigenvalue in the case of an intersite Maxwell front.
\end{lemma}

\begin{proof}
Consider the eigenvalue problem for the operator $L_{+}$:
\begin{equation}\label{eqn:L_{+}_eigval_eqn}
    L_{+} w = \alpha w.
\end{equation}
For the onsite and intersite fronts ($n_{0} = 0$ or $n_{0} = 1/2$), the exponentially small growing term in the remainder vanishes. Consequently, $\tilde{\phi}_{M} \to 0$ as $\tilde{x} \to -\infty$ and $\tilde{\phi}_{M} \to 2/3$ as $\tilde{x} \to \infty$ exponentially fast. This leads to $F_{\phi}(\tilde{\phi}_{M};\mu_{M}) \to -\mu_{M}$ as $|\tilde{x}| \to \infty$. A far-field analysis of $L_{+}$, using Weyl's Essential Spectrum Lemma, shows that the essential spectrum of $L_{+}$ is bounded below by $-\mu_{M} = 2/9$.

Next, consider the expansion of $\Delta$ in powers of $\varepsilon$. The operator $L_{+}$ can be expressed as
\begin{align}
    L_{+} = L_{+}(\varepsilon) = -2 \sum_{m \geq 1} \frac{\varepsilon^{2m-2}}{(2m)!} \partial_{x}^{2m} + F_{\phi}(\tilde{\phi}_{M},\mu_{M}), \ \ \text{as } \varepsilon \to 0.
\end{align}
Since $\tilde{\phi}_{M} \to \phi_{0}$ as $\varepsilon \to 0$, $L_{+}$ simplifies to the differential operator
\begin{align}
    L_{+}(0) = -\partial_{x}^{2} + F_{\phi}(\phi_{0};\mu_{M}).
\end{align}
Since the solution $\phi_{0}(\tilde{x})$ is translationally invariant (as defined in \eqref{eqn:phi_0_equation}), the kernel of $L_{+}(0)$ is spanned by $\phi_{0}'(\tilde{x})$. Noting that $\phi_{0}'(\tilde{x})$ is strictly positive, we conclude that the discrete spectrum of $L_{+}(0)$ contains no negative eigenvalues.

We now analyze the perturbation of the zero eigenvalue when $\varepsilon \neq 0$. To do so, we expand $w$ in an asymptotic power series in $\alpha$ as
\begin{equation}
    w = w_{0} + \alpha w_{1} + \dots, \quad (\alpha \ll 1).
\end{equation}
At leading order, this gives the equation
\begin{equation}
    L_{+}(\varepsilon)w_{0} = 0.
\end{equation}

To analyze the bifurcation of the zero eigenvalue as \( 0 < \varepsilon \ll 1 \), we perturb around the translationally invariant eigenfunction in the \( \varepsilon = 0 \) limit. Specifically, we consider
\begin{equation}
    w_{0} = \partial_{\tilde{x}}\tilde{\phi}_{M}.
\end{equation}
Considering $\tilde{\phi}_{M}$ as
\begin{equation}
    \phi_{M}(\tilde{x}) = \phi_{0}(\tilde{x}) + 2\pi \varepsilon^{-4}e^{-3\sqrt{2}\pi^{2}/\varepsilon}|\Lambda_{1}|\left[-\text{Re }G(\tilde{x})\sin(2\pi\tilde{x}/\varepsilon) + \text{Im }G(\tilde{x}) \cos(2\pi\tilde{x}/\varepsilon)\right] \ \ (0 < \varepsilon \ll 1),
\end{equation}
we observe that when $0 < \varepsilon \ll 1$, $w_{0}$ exhibits an exponentially growing tail as $\tilde{x}\to \infty$
\begin{equation}
    w_{0} \sim -\frac{27}{2}\pi^{2} \varepsilon^{-5} e^{-3\sqrt{2}\pi^{2}/\varepsilon} |\Lambda_{1}|\cos(2\pi n_{0}) e^{\sqrt{2}\tilde{x}/3}.
\end{equation}
This growing tail will be counterbalanced by the solution at $\mathcal{O}(\alpha)$. At this order, we have
\begin{equation}
    L_{+}(\varepsilon)w_{1} = w_{0}.
\end{equation}

As $\varepsilon \to 0$, $\tilde{\phi}_{M} \sim \phi_{0}$ and $w_{0} \sim \phi_{0}'(\tilde{x})$. Thus, as $\varepsilon \to 0$, we asymptotically obtain
\begin{equation}
    L_{+}(0)w_{1} \sim \phi_{0}'(\tilde{x}).
\end{equation}
Using a reduction of order, we express $w_{1}$ as $w_{1} = v(\tilde{x})\phi_{0}'(\tilde{x})$, where $v(\tilde{x})$ satisfies
\begin{equation}
    v'(\tilde{x}) = -\frac{1}{[\phi_{0}'(\tilde{x})]^{2}} \int_{-\infty}^{\tilde{x}} [\phi_{0}'(s)]^{2} \, ds.
\end{equation}
As $\tilde{x} \to \infty$, this yields
\begin{equation}
    v'(\tilde{x}) \sim -\frac{1}{2\sqrt{2}}e^{2\sqrt{2}\tilde{x}/3},
\end{equation}
and $w_{1}$ develops a growing tail:
\begin{equation}
    w_{1} \sim -\frac{1}{6\sqrt{2}}e^{\sqrt{2}\tilde{x}/3}, \quad \tilde{x} \to \infty.
\end{equation}
Combining the contributions of $w_{0}$ and $w_{1}$, we have
\begin{equation}
    w \sim w_{0} + \alpha w_{1} \sim \left[-\frac{27}{2}\pi^{2} \varepsilon^{-5} e^{-3\sqrt{2}\pi^{2}/\varepsilon} |\Lambda_{1}|\cos(2\pi n_{0}) - \frac{\alpha}{6\sqrt{2}}\right]e^{\sqrt{2}\tilde{x}/3}, \quad \text{as } \varepsilon \to 0, \; \tilde{x} \to \infty.
\end{equation}
Eliminating the growing tail, we find the asymptotic formula for $\alpha$:
\begin{equation}
    \alpha \sim -81\sqrt{2}\pi^{2} \varepsilon^{-5} e^{-3\sqrt{2}\pi^{2}/\varepsilon} |\Lambda_{1}|\cos(2\pi n_{0}), \quad \text{as } \varepsilon \to 0.
\end{equation}

From this, we conclude that as $\varepsilon \to 0$, $L_{+}(\varepsilon)$ has one exponentially small eigenvalue. Specifically, for $n_{0} = 0$ (onsite fronts), $L_{+}$ has a single negative eigenvalue, while for $n_{0} = 1/2$ (intersite fronts), $L_{+}$ has a single positive eigenvalue.\qed
\end{proof}

Now, we invoke a theorem of Chugunova and Pelinovsky \cite{chugunova2006counteigenvaluesgeneralizedeigenvalue} in counting the eigenvalues of the spectral problem \eqref{eqn:eigval_eqn} to obtain the following result.

\begin{theorem} \label{thm:eigval_count}
Suppose that $L_{\pm}$ have trivial kernels in $l^{2}(\Z)$, and that $L_{\pm}$ possess $n_{\pm}$ negative eigenvalues. Further, assume that all embedded eigenvalues of the spectral problem \eqref{eqn:eigval_eqn} are algebraically simple. Under these conditions, the spectral problem \eqref{eqn:eigval_eqn} exhibits $N_{c}$ complex eigenvalues in the first quadrant, $N_{i}^{-}$ purely imaginary eigenvalues with positive imaginary part satisfying $\langle v, L_{+}^{-1}v \rangle \leq 0$, $N_{r}^{+}$ real positive eigenvalues for which $\langle v, L_{+}^{-1}v \rangle \geq 0$, and $N_{r}^{-}$ real positive eigenvalues satisfying $\langle v, L_{+}^{-1}v \rangle \leq 0$. These eigenvalue counts are related through the equations:
\begin{align}
    N_{r}^{-} + N_{i}^{-} + N_{c} &= n_{+}, \label{eqn:count_eigval_1}\\
    N_{r}^{+} + N_{i}^{-} + N_{c} &= n_{-}. \label{eqn:count_eigval_2}
\end{align}
\end{theorem}

\begin{proof}
    Since $\tilde{\phi}_{M} \to 0$ as $\tilde{x} \to -\infty$ and $\tilde{\phi}_{M} \to 2/3$ as $\tilde{x} \to \infty$, both at an exponentially fast rate, it follows that $F_{\phi}(\tilde{\phi}_{M};\mu_{M}) \to -\mu_{M}$ exponentially fast as $|x| \to \infty$. Consequently, by Weyl's Essential Spectrum Lemma, the essential spectra of $L_{+}$ and $L_{-}$ are respectively bounded below by $-\mu_{M} = 2/9$ and $0$. 

Now, since the kernel of $L_{+}$ is empty in $l^{2}(\Z)$, we may consider the eigenvalue equation \eqref{eqn:eigval_eqn_LmLp} as the generalized eigenvalue equation
\begin{equation} \label{eqn:eigval_eqn_2}
    L_{-}u = \gamma L_{+}^{-1} u, \quad \gamma = -\lambda^{2}.
\end{equation}
Furthermore, we can shift this equation, rewriting it as
\begin{equation} \label{eqn:eigval_eqn_3}
    \left(L_{-} + \delta L_{+}^{-1}\right)u = (\gamma + \delta) L_{+}^{-1} u,
\end{equation}
for sufficiently small $\delta > 0$.

Since the properties P1 and P2 from \cite{chugunova2006counteigenvaluesgeneralizedeigenvalue} are satisfied by $L_{\pm}$, Theorem 3 of \cite{chugunova2006counteigenvaluesgeneralizedeigenvalue} applies. As a result, we obtain the relations
\begin{align}
    N_{r}^{-} + N_{i}^{-} + N_{c} &= \dim\left(\mathcal{H}^{-}_{L_{+}^{-1}}\right), \\
    N_{r}^{+} + N_{i}^{-} + N_{c} &= \dim\left(\mathcal{H}^{-}_{L_{-}+\delta L_{+}^{-1}}\right),
\end{align}
where $\mathcal{H}^{-}_{A}$ denotes the negative invariant subspaces of $l^{2}(\Z)$ associated with the operator $A$. It then follows that $\dim(\mathcal{H}_{L_{+}^{-1}}) = n_{+}$.

Finally, we need to show that $\dim(\mathcal{H}_{L_{-}+\delta L_{+}^{-1}}) = \dim(\mathcal{H}_{L_{-}}) = n_{-}$. This can be demonstrated as follows:
    
    The inequality $\dim(\mathcal{H}_{L_{-}}) \leq \dim(\mathcal{H}_{L_{-} + \delta L_{+}^{-1}})$ follows from the continuity of the eigenvalues and the relative compactness of $L_{+}^{-1}$ with respect to $L_{-}$. The key issue is whether, for small $\delta > 0$, an edge bifurcation causes an eigenvalue of $L_{-} + \delta L_{+}^{-1}$ to emerge from the lower edge of the essential spectrum of $L_{-}$. To investigate this, consider the eigenvalue problem
\begin{equation}\label{eqn:eigval_eqn_Lm+dLp-1}
    (L_{-} + \delta L_{+}^{-1})w = \omega w.
\end{equation}
This equation can also be written as
\begin{equation}
    (L + \delta B)w = \omega w,
\end{equation}
where
\begin{align}
    L &= L_{-} + \delta L_{+,\infty}^{-1}, \, 
    B = L_{+}^{-1}\left(L_{+,\infty} - L_{+}\right)L_{+,\infty}^{-1} = L_{+}^{-1}\left(2|\tilde{\phi}_{M}| - 3|\tilde{\phi}_{M}|^{2}\right)L_{+,\infty}^{-1}.
\end{align}
Here, $L_{+,\infty} = -\dfrac{1}{\varepsilon^{2}}\Delta - \mu_{M}$ represents the limiting form of the operator $L_{+}$ as $|x| \to \infty$. The operator $B$ constitutes a relatively compact perturbation to $L$, and the essential spectrum of $L$ is bounded below by $\frac{\delta}{-\mu_{M}} > 0$. 

By applying the theory of edge bifurcations \cite{kapitula_stanstede2004_eigenvalues_evans}, any edge bifurcation that occurs when $\delta \neq 0$ and originates at the lower edge of the essential spectrum of $L$ would appear as
\begin{equation}
    \omega(\delta) = \frac{\delta}{-\mu_{M}} - a \delta^{2} + \mathcal{O}(\delta^{3}),
\end{equation}
where $a > 0$ is a constant. For sufficiently small $\delta > 0$, $\omega(\delta) > 0$, which implies that the number of negative eigenvalues of $L_{-} + \delta L_{+}^{-1}$ remains unchanged. Consequently, $\dim(\mathcal{H}^{-}_{L_{-}+\delta L_{+}^{-1}}) = \dim(\mathcal{H}^{-}_{L_{-}}) = n_{-}$ for small $\delta > 0$.

Thus, we have established that \eqref{eqn:count_eigval_1} and \eqref{eqn:count_eigval_2} hold.\qed
\end{proof}

In our case, we have established that $L_{+}$ has a trivial kernel in $l^{2}(\Z)$. This is because, although $L_{+}\partial_{x_{0}}\tilde{\phi}_{M} = 0$, the function $\partial_{x_{0}}\tilde{\phi}_{M}$ has an exponentially growing tail and thus does not belong to $l^{2}(\Z)$. Moreover, $L_{-}\tilde{\phi}_{M} = 0$ and $\tilde{\phi}_{M} \in l^{\infty}(\Z)$ but not in $l^{2}(\Z)$, thus the kernel of $L_{-}$ is also empty in $l^{2}(\Z)$. As a result, we can directly apply the results of Theorem \ref{thm:eigval_count}. Furthermore, since $L_{-}\tilde{\phi}_{M} = 0$ and $\tilde{\phi}_{M}(n)$ is strictly positive for all $n \in \Z$, Sturm-Liouville theory guarantees that $n_{-} = 0$. This immediately implies that $N_{c} = N_{i}^{-} = N_{r}^{+} = 0$.

To determine the number of real positive eigenvalues, we apply the first equality \eqref{eqn:count_eigval_1}:
\begin{align}
    N_{r}^{-} = n_{+}.
\end{align}

Using Lemma \ref{lem:eigval_Lp} and Theorem \ref{thm:eigval_count}, we find that for the onsite Maxwell front, $N_{r}^{-} = 1$. This indicates that the onsite Maxwell front is unstable due to the presence of one real positive eigenvalue. In contrast, the intersite Maxwell front is stable since, in that case, $n_{+} = 0$, which leads to $N_{r}^{-} = 0$. Similar analysis yields similar stability results for the two solutions when $0 < |\delta\mu| < \max |\delta\mu|$.

\section{Conclusion}\label{sec5}

We have comprehensively studied the formation, stability, and dynamical behavior of droplets and bubbles in the quadratic-cubic discrete NLS equation. Using exponential asymptotic methods, our analysis identified a pinning region, a parameter interval where multiple stable states coexist and are interconnected through homoclinic snaking. This snaking behavior manifests as a series of intertwined solution branches in the bifurcation diagram.

We established a precise exponential relationship between the coupling strength and the width of the pinning region. We demonstrated that the pinning region narrows exponentially as the coupling strength increases. Furthermore, we showed that intersite Maxwell fronts remain stable while onsite fronts are inherently unstable. We validated our analysis with numerical simulations and obtained a good agreement. Finally, our detailed study on the asymptotics of the critical eigenvalue of the linear eigenvalue problem, which governs the stability of these structures, will be reported in a separate work.

\section*{Data availability}
No data was used for the research described in the article.

\section*{Declaration of competing interest} 
The authors declare that they have no known competing financial interests or personal relationships that could have appeared to influence the work reported in this paper.

\section*{CRediT authorship contribution statement}
The manuscript was written with contributions from all authors. All authors have given their approval to the final version of the manuscript.\\

\textbf{FTA}: Formal Analysis, Investigation, Software, Validation, Writing - Original Draft, Writing - Review \& Editing; \textbf{HS}: Conceptualization, Methodology, Validation, Supervision, Writing - Original Draft, Writing - Review \& Editing.

\section*{Acknowledgements}
The authors thank the referee for their careful reading. HS acknowledged support by Khalifa University through a Competitive Internal Research Awards Grant (No.\ 8474000413/CIRA-2021-065) and Research \& Innovation Grants (No.\ 8474000617/RIG-S-2023-031 and No.\ 8474000789/RIG-S-2024-070).

	\bibliographystyle{elsarticle-num}
	\bibliography{Report_dnls23Notes}
	
\end{document}